\documentclass[5p]{elsarticle}

\usepackage{amsmath}
\usepackage{graphicx}
\usepackage{xcolor}
\usepackage{epstopdf}
\usepackage{natbib}
\usepackage{xfrac}
\usepackage[normalem]{ulem}
\usepackage{multirow}
\usepackage{physics}
\usepackage{amssymb}
\usepackage{verbatim}

\usepackage{mathtools, cuted}
\usepackage[english]{babel}
\usepackage{lipsum}

\begin{document}

\title{Numerical solver for the time-dependent far-from-equilibrium Boltzmann equation}

\author[1,2]{M.~Wais}
\author[2]{K. Held}
\author[1]{M. Battiato\corref{cor1}}
\ead{marco.battitao@ntu.edu.sg}

\cortext[cor1]{Corresponding author}
\address[1]{Nanyang Technological University, 21 Nanyang Link, Singapore, Singapore}
\address[2]{Institute of Solid State Physics, TU Wien,  Vienna, Austria}



\begin{abstract}
The study of strongly out-of-equilibrium states and their time evolution towards thermalization is critical to the understanding of an ever widening range of physical processes. 
%
We present a numerical method that for the first time allows for the numerical solution of the most difficult part of the time-dependent Boltzmann equation: the full scattering term. Any number of bands (and quasiparticles) with arbitrary dispersion, any number of high order scattering channels (we show here four legs scatterings: electron-electron scattering) can be treated far from equilibrium. No assumptions are done on the population and all the Pauli-blocking factors are included in the phase-space term of the scattering. The method can be straightforwardly interfaced to a deterministic solver for the transport. Finally and most critically, the method conserves to machine precision the particle number, momentum and energy for any resolution, making the computation of the time evolution till complete thermalization possible. 

We apply this approach to two examples, a metal and a semiconductor, undergoing thermalization from a strongly out-of-equilibrium laser excitation. These two cases, which are in literature treated hitherto under a number of approximations, can be addressed free from those approximations and straightforwardly within the same numerical method.
\end{abstract}

\begin{highlights}
\item Implementation of Boltzmann equation with full two-particle scattering
\item Efficient calculation and storage of the scattering tensor
\item Exact energy, particle number and momentum conservation in time-propagation
\end{highlights}


\maketitle



\section{Introduction}
The development of femtosecond lasers has granted access to ultrafast dynamics in solids ranging from femtoseconds to picoseconds
\cite{Fujimoto1984,Elsayed1987,Schoenlein1987,Brorson1990,Fann1992,Hertel1996,Beaurepaire1996,Fatti2000,Stamm2007}
. These timescales are particularly complicated to describe as not only the system is far from equilibrium, but also exhibits important changes as it undergoes thermalization. It is therefore crucial to precisely describe both the  far-from-equilibrium state (as it directly controls effects like transport) and its time evolution under many-body effects.
What makes the description even more challenging is that a range of effects arise due to the complex interplay of several degrees of freedom, and a detailed description of the material, or the concomitant treatment of both scatterings and transport is necessary to describe them \cite{Malinowski2008, Battiato2010,Kampfrath2013, Battiato2016,Freyse2018}. Moreover typical femtosecond laser excitations can produce very strong excitation regimes shifting the chemical potential significantly and even triggering dynamic metal to insulator transitions \cite{Terada2008,Crepaldi2012,Garcia2011}. This dramatically affects the scattering lifetimes and brings into the picture a number of higher order scatterings which can often be neglected close to equilibrium. 

One of the most effective theoretical frameworks to calculate non-equilibrium dynamics is the Boltzmann equation (BE)
\cite{Snoke2011,Rethfield2002,Vatsal2017}, as it allows for an extremely precise treatment of both scattering (as long as proper quasiparticles are used) and transport, including many quantum mechanical effects (but not interference). 
The Boltzmann equation, originally invented for the kinetic theory of gases \cite{Boltzmann1872}, is widely used in different research fields: all the way from nuclear physics \cite{Abe1996} to plasma- and astrophysics \cite{Colonna2016,Rockwood1973,Li1998,Maruca_2012,Brizard1995} or cosmology \cite{Dodelson2014,Ferreira1998}, from fluid mechanics \cite{Chen1998} to traffic simulations \cite{Helbing2001}. In solid state physics it was employed from the outset \cite{Lorentz1905,Sommerfeld1928}, and is applied nowadays from semiconductors \cite{Semiconductor1969,Abdallah1996,Choquet2000,Majorana2004,Caceres2006} to perovskite solar cells \cite{Motta2015, Wais2018}, from thermal transport \cite{Cahill2003,Paothep2010,Madsen2006} to ultrafast dynamics \cite{Battiato2016,Tani2012,Battiato2020,Cheng2019}.

The BE has two main parts: transport terms and the collision integral. A number of methods have been successfully used for the transport part. On the other hand the scattering term still poses a number of critical challenges. We will only address the scattering integral in the present work (but the numerical method proposed has been developed with the important constraint to be fully compatible with highly performing numerical methods for the transport part).

Addressing the time propagation of the BE scattering term requires two steps: All quasiparticles dispersions and all the scatterings matrix element have to be calculated. This first step requires a wide range of techniques and poses a variety of challenges. A vast literature is present on the topic, and we will not address that in this work. In the following we will assume that the matrix elements have been either calculated or satisfactorily approximated for all the scatterings one wants to include in the BE.

The second step is the actual time propagation of the scattering integral \cite{Mahan2000,Snoke2011,Snoke2007}.
 This step has been already addressed in the presence of a number of approximations which have proven to be very successful  for several applications~\cite{Rethfield2002,Battiato2010,Battiato2016,Maldonado2017,Vatsal2017,Sadasivam2017,Sanchez2017,Wais2018,Battiato2018}.
  However the numerical solution of the full scattering term poses a number of hitherto unresolved challenges \cite{Kurosawa1971,Jacoboni1983,Fischetti1988,Medvedev2011,Mattei2017,Nenno2018}.

The first issue is the exact and concomitant conservation of particle number, momentum and energy. While ensuring the first one is easy, the real challenge is the exact conservation of both momentum and energy simultaneously. While this might seem a rather technical point, breaking those conservations imposes strong limitations to the applicability of a numerical method, as violating some of those conservations  makes the numerical method ill-suited for medium or long time propagation. A small error in, for instance, the total energy will pile up over every timestep. The numerical time propagation will either (in the fortunate cases) thermalise to a wrong thermal equilibrium, or most probably never thermalise (for instance leading the system to heat up indefinitely). The implication is that any numerical method aiming at describing the full range of timescales and the thermalisation of a closed system must exactly satisfy those conservation laws even in the presence of discretisation error. Notice that in the time propagation of an open system (i.e.~one that can exchange momentum and/or energy with a bath, or where one of the populations have been approximated by a fixed distribution), on the other hand, a numerical method that does violate those conservations, but only by an amount proportional to the computational error, can still provide accurate and stable solutions. This is because for an open system, the steady state will be completely defined by the bath and not by long-time propagations of small errors. All numerical methods that address approximated versions of the BE scattering, benefit from this partial insensitivity to this problem. Notice that describing the full thermalisation of a material driven strongly out-of-equilibrium eventually requires it to be treated as a closed system, i.e.~all the quasiparticles populations should be kept free to be modified by the scatterings and not kept fixed (or equivalently treated as a bath).
In this work we will show a numerical method that, by solving the problem of exact and concomitant conservation laws, finally allows for the time propagation of the full BE scattering term for closed systems without any approximation. 

The mathematical structure of the scattering integral remains the same, regardless of the scattering type. It depends only on the number of states (which we will call legs, and which can belong to even different kinds of quasiparticles) involved in the scattering. Therefore the numerical approach to the time propagation of the scattering we develop here is general to any kind of scattering. However there is a challenge that can become crippling depending on the number of legs of the scatterings included: the scaling of the numerical cost with precision. In case of a three legs scattering (for instance electron-phonon scattering) in general the full scattering integral depends on three unknown independent populations. This makes the scattering operator cubic which is equivalent to a 4 dimensional scattering tensor in a basis, as we will see below. Each of its dimensions grows with the precision used to describe the population in the basis (for instance that could be the number $N$ of $k$ points in a finite-differences scheme). The entries of a 4 dimensional tensor would be $N^4$ (these are usually compressed, but the cost still remains large). In case of a 4 legs scattering (for instance electron-electron scattering) the cost becomes a further $N$ times larger than for a 3 legs scattering, which can easily render the whole problem practically unsolvable. Notice that this issue is strongly mitigated if some of the populations are kept constant. We here also address the crippling problem of the cost scaling.

The outline of the paper is as follows: In Sec.~\ref{chap:fokker} we introduce the Boltzmann-scattering term for electron-electron collisions and highlight its relation to the quantum Fokker-Planck equation. In Sec.~\ref{chap:scratesmain} we explain how the scattering rate of an electron, added to the equilibrium system, is calculated within the Boltzmann framework.

Sec.~\ref{chap:deiscretization} is dedicated to the explanation of the novel, numerical method for the calculation of the Boltzmann-scattering term. In particular, in Sec.~\ref{chap:discrBasis} we introduce the basis we use and project the scattering term onto it, leading to the so-called scattering-tensor. In Sec.~\ref{chap:calcSC} we discuss how to actually calculate the tensor and how the conservation of extensive thermodynamic quantities results in symmetries in the scattering tensor. In Sec.~\ref{chap:timeprop} we quickly discuss how the time-propagation is done and in Sec.~\ref{chap:scrates} we present a way to calculate equilibrium scattering rates exploiting the scattering-tensor. 

In Sec.~\ref{chap:nonequdyn} we apply our method to two 2-band systems, one metal and one semiconductor, with different initial non-equilibrium electron populations. Firstly, we calculate the equilibrium scattering rates for both system for all possible scattering processes (Sec.~\ref{chap:scatteringRatesTest}). 
Then, we discuss the thermalization dynamics for an initial distribution that is purely energy dependent and particle-hole symmetric for the metallic- (Sec.~\ref{chap:partholesymdel0}) and the semiconducting- (Sec.~\ref{chap:partholesymdel1}) case. This is followed by the thermalization dynamics for an initial distribution that is explicitly momentum-dependent without particle-hole symmetry, again for the metal (Sec.~\ref{chap:partholenonsymdel0}) and the semiconductor (Sec.~\ref{chap:partholenonsymdel1}). 

Finally, Sec.~\ref{chap:conclusion} summarizes our results.

\section{Time evolution in the presence of scatterings}\label{chap:timeevo}

\subsection{The scattering integral: the quantum Fokker-Planck equation}\label{chap:fokker}

The Boltzmann equation is composed of a transport part and a collision term. The most challenging part is the second, often referred to as scattering integral also known as Boltzmann collision integral. 

When transport is not relevant, the Boltzmann equation becomes equivalent to the quantum Fokker-Planck equation\cite{Snoke2007}. It provides the time derivative of the expectation value of the number operator of a given state, in the presence of an interaction term, calculated within first-order time-dependent perturbation theory for a system with a hamiltonian of the type
\begin{equation}
 	H= \sum_{n, \bold k} \epsilon_n\left(\bold k\right) c^\dagger_{\underset{n}{\bold k}} c_{\underset{n}{\bold k}} +\frac{1}{2} \sum_{\underset{n_0,n_1,n_2,n_3}{\bold k_0, \bold k_1, \bold k_2,\bold k_3}} V_{\underset{n_0,n_1,n_2,n_3}{\bold k_0, \bold k_1, \bold k_2,\bold k_3}} c^\dagger_{\underset{n_2}{\bold k_2}}c^\dagger_{\underset{n_3}{\bold k_3}} c_{\underset{n_1}{\bold k_1}}  c_{\underset{n_0}{\bold k_0}} \label{eq:initialhamilton}
\end{equation}
where $\bold k$ is the crystal momentum. The index $n$ contains quasiparticle type, band number, as well as other applicable quantum numbers like, for instance, spin. Furthermore, $\epsilon_n\left(\bold k\right)$ is the quasi-particles dispersion and $V_{\underset{n_0,n_1,n_2,n_3}{\bold k_0, \bold k_1, \bold k_2,\bold k_3}}$ is the matrix-transition-element of the interaction potential with two two-particle states 

($V_{\underset{n_0,n_1,n_2,n_3}{\bold k_0, \bold k_1, \bold k_2,\bold k_3}} = \bra{n_2\bold k_2,n_3\bold k_3} \hat V \ket{n_0\bold k_0,n_1 \bold k_1}$).
 The Hamiltonian above describes the evolution of interacting quasi-particles in a solid, and can be easily generalized to include different types of quasiparticles and scatterings.

The time dependence of the expectation value of the number operator $\left< N_{\underset{n}{\bold k}} \right> = \left< c^\dagger_{\underset{n}{\bold k}}c_{\underset{n}{\bold k}} \right>$ at a time $t$ is customarily written as $f_{n}(t,\bold k)$ and called distribution function in the context of the  Boltzmann equation. For instance, in presence of weak electron-electron (in quasiparticle sense) interaction (necessary for the applicability of the first-order time-dependent perturbation theory), where an electron of band $n_0$ scatters with an electron of band $n_1$ to end up in the bands $n_2$ and $n_3$ (and its time-reversed process; $n_0 + n_1 \leftrightarrow n_2 + n_3$), the Boltzmann collision integral (aka the quantum Fokker-Planck equation) for the time derivative of the distribution-function in band $n_0$ reads \cite{Snoke2007}
\begin{strip}
\begin{equation}
\begin{split}
\left ( \frac{\partial f_0}{\partial t} \right )_{n_0 + n_1 \leftrightarrow n_2 + n_3 } & =~ \left ( 1- \frac{1}{2}\delta_{n_2,n_3} \right ) \sum_{\bold G}  \iiint_{{V_{BZ}}^3}   \mathrm d^d \bold k_1 \; \mathrm d^d \bold k_2 \; \mathrm d^d \bold k_3   ~ w_{0123}^\textrm{e-e}~ \delta(\bold k_0 +\bold k_1 - \bold k_2 - \bold k_3+ \bold G) \\
&\times  \delta \left ( \epsilon_{n_0}(\bold k_0) + \epsilon_{n_1}(\bold k_1) - \epsilon_{n_2}(\bold k_2) -\epsilon_{n_3}(\bold k_3) \right ) \underbrace{ \left [ (1-f_0)(1-f_1)f_2f_3 - f_0 f_1 (1-f_2)(1-f_3) \right ] }_{\equiv \mathcal P_{0123}} \textrm{ .} \label{eq:eecollisionScatNum}
\end{split} 
\end{equation}
\end{strip}
%
Here we have used the shorthand notation $f_i \equiv f_{n_i}(t,\bold k_i)$, $V_{BZ}$ is the Brillouin zone volume, $d$ is the spatial dimension of the system described and $\mathcal P_{0123} = \mathcal P_{0123} (t, \bold k_0,\bold k_1, \bold k_2 ,\bold k_3)$ is the so-called phase-space factor. The symbol $\sum_{\bold G}$ is the sum over all reciprocal lattice vectors $\bold G$ and accounts for Umklapp scatterings. The first Dirac delta enforces momentum conservation, while the second one enforces energy conservation. The subscript of the time derivative on the lefthand side indicates the  scattering that is driving the change in population. 
Starting from the Hamiltonian given by Eq.~\eqref{eq:initialhamilton} the collision integral Eq.~\eqref{eq:eecollisionScatNum} can be derived in first-order time-dependent perturbation theory (Fermi's Golden rule)
. Physically this means that phase coherence between scatters is neglected;  occupations but not phases are encoded in $f_i$. This is the underlying approximation of the Boltzmann equation (BE). Often further approximations are done such as a relaxation time approximation for Eq.~\eqref{eq:eecollisionScatNum}. In our paper, we show how to treat the collision integral and hence the BE exactly.
In the presence of several scattering channels, one simply sums up the time derivatives due to each scattering channel individually. The expression in Eq.~\eqref{eq:eecollisionScatNum} can be easily generalized to different types of scatterings, for instance between different quasiparticles, or involving a different number of states, and to different kinds of quaiparticles. The scattering amplitude $w_{0123}^\textrm{e-e}$ is proportional to the transition matrix element squared and it depends, in general, on all the involved states, through their momenta $\bold k_i$ and band numbers $n_i$,
\begin{equation}
w_{0123}^\textrm{e-e} = \frac{V^2}{{(2 \pi)}^{2d - 1}} \left | \bra{n_2\bold k_2,n_3\bold k_3} \hat V \ket{n_0\bold k_0,n_1 \bold k_1} \right | ^2 \textrm{ .}
\end{equation}
The factor $\left ( 1- \frac{1}{2}\delta_{n_2,n_3} \right )$ in Eq.~\eqref{eq:eecollisionScatNum} is needed to prevent double counting and will be absorbed into $w_{0123}^\textrm{e-e}$ in the following for brevity.

Let us now highlight an important requirement to abide to, before using Eq.~\eqref{eq:eecollisionScatNum}. The Fokker-Planck equation is derived within first-order time-dependent perturbation theory. Therefore its applicability is limited to cases where the interaction term can be considered a perturbation. This is almost never the case for bare particles. It is therefore necessary to rewrite the hamiltonian in terms of weakly interacting quasiparticles, and write the Boltzmann equation for those. For instance, if the system displays strongly bound excitons, it is them that have to enter the Boltzmann equation, and not the electron and the hole individually. 

The numerical treatment of Eq.~\eqref{eq:eecollisionScatNum} in realistic band structures presents  several difficulties. (i) The integral is $3d$-dimensional, resulting in an $6$-dimensional integral in $2D$, and 9-dimensional in $3D$. (ii) The integral is a quartic operator on the distribution function resulting in bad scaling with respect to the number of bands and momentum patches. (iii) The integral contains two delta distributions, one of which has a highly non trivial argument. The delta that ensures momentum conservation depends on the integration variables (i.e. the momenta) linearly, hence, it can be inverted analytically. However, the delta distribution for energy conservation depends on the momenta through the dispersion relations. In general, they have arbitrary shapes and the delta distribution describes a highly complex $(3d-1)$-dimensional hypersurface in the 3$d$-dimensional integration domain. iv) The analytic expression of the scattering integral conserves quantities like number of particles, momentum (only up to a reciprocal lattice vector) and energy exactly. If not treated carefully, numerical errors ensuing from the integration of the scattering term in general lead to a breaking of these conservation laws. This is not a problem in steady state calculations, but it becomes critical in time evolution, where these errors pile up at every timestep leading quickly to completely unphysical and meaningless results. It is in principle not even guaranteed that the numerical solution will converge to a steady state (i.e.~thermalize), as expected in reality.

\subsection{Scattering rates} \label{chap:scratesmain}

Besides the time propagation, other very interesting quantities can be calculated from the scattering integral. The most interesting are the $\bold k$-resolved scattering rates, which are the inverse of the lifetimes.

It is easy to prove that if the distribution functions for all the particles in the system are Fermi-Dirac for fermions, and Bose-Einstein for bosons, then the scattering integral gives a zero time derivative. If we modify the density, the scattering integral will drive back the system to equilibrium. When the modification (either positive or negative) is small and very localised in $\bold k$, one can prove that the decay back to equilibrium is a simple exponential decay, with a time constant that is the $\bold k$-resolved lifetime. It is often more convenient to work in terms of the inverse lifetimes, the $\bold k$-resolved scattering rate. 

The scattering rates $\lambda_{n_0} (\bold k)$ of small, $\bold k$-localised deviations from equilibrium in the population of a band $n_0$ due to a given scattering provide a rich amount of information on that scattering, and are therefore very helpful in the physical interpretation of the dynamics. 
It can be shown (Appendix~\ref{chap:Ascatteringrates}) that the scattering rates $\lambda_{n_0} (\bold k)$ can be written as (again for the specific case of electron-electron scattering):

\begin{strip}
\begin{equation}
\begin{split}
&\Big( \lambda_{n_0} (\bold k_0) \Big)_{n_0 + n_1 \leftrightarrow n_2 + n_3 }  =~  \sum_{\bold G}  \iiint_{{V_{BZ}}^3}   \mathrm d^d \bold k_1 \; \mathrm d^d \bold k_2 \; \mathrm d^d \bold k_3   ~w_{0123}^\textrm{e-e}~\delta(\bold k_0 +\bold k_1 - \bold k_2 - \bold k_3+ \bold G) \times\\
&\;\;\;\; \times  \delta \left ( \epsilon_{n_0}(\bold k_0) + \epsilon_{n_1}(\bold k_1) - \epsilon_{n_2}(\bold k_2) -\epsilon_{n_3}(\bold k_3) \right ) \left [ (1-f_{eq,1})f_{eq,2}f_{eq,3} +  f_{eq,1} (1-f_{eq,2})(1-f_{eq,3}) \right ]  \textrm{ ,} \label{eq:eecollisionScatRate6Analytic}
\end{split} 
\end{equation}
\end{strip}
where $f_\textrm{eq}$ is the appropriate equilibrium distribution for that particle type.

\begin{figure*}
 \centering
 \includegraphics[width=1\textwidth]{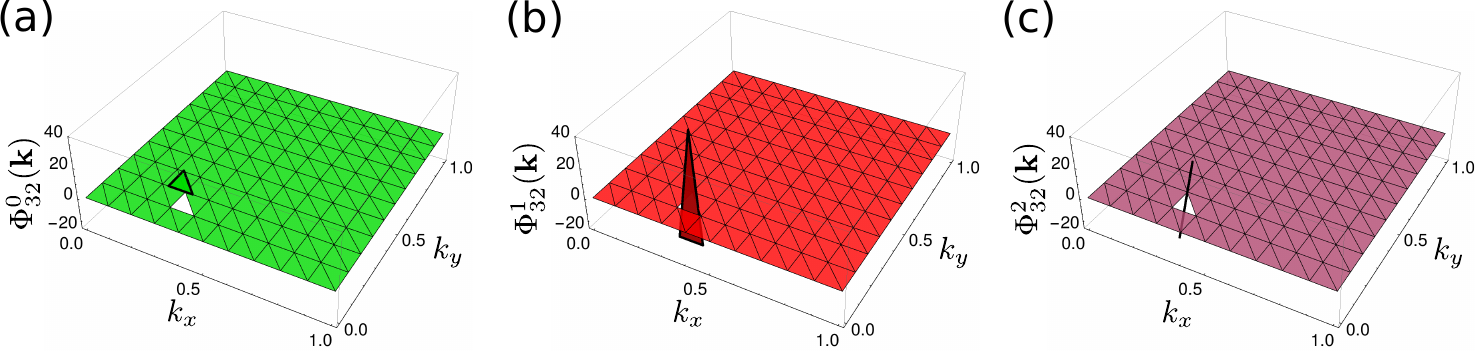}
\caption{The three basis functions for the element $I=32$ as defined by Eq.~\eqref{eq:dgBasis2D}; (a) $\Phi_{32}^0(\cdot)$, (b) $\Phi_{32}^1(\cdot)$, (c) $\Phi_{32}^2(\cdot)$.
}
  \label{fig:basisfunmed}
\end{figure*}

\section{Discretisation method}\label{chap:deiscretization}

\subsection{Semidiscretised problem and scattering tensor}\label{chap:discrBasis}

We will address the scattering problem numerically using a spectral approach, i.e.~we will project the collision integrals as well as the distribution functions onto a basis. In the following we will refer explicitly to the $2D$ case, but the method is general to any dimension. The success of a spectral method lies in an optimal choice of basis function. In order to construct the basis functions, we first construct an irregular triangular grid (a tetrahedral grid in $3D$), also called mesh, of an appropriately chosen domain for each band $n$. Notice that the domain for each band does not need to cover the whole Brillouin zone: this leads to major computational savings in cases of highly dispersive bands, where large parts of the band extends onto uninteresting energy regions, which can be easily excluded from the computation. Moreover every band can have different meshes. Finally notice that in general the domain can be irregular and the mesh can be more refined in regions where a higher resolution is required. Such kind of meshes are routinely used in numerical methods like finite elements and the triangles are often referred to as elements.

In the numerical part of this work we have used structured meshes (see Fig.~\ref{fig:basisfunmed}) which are the same for both bands. However let us stress  that the approach is valid for unstructured meshes defined independently for each band. 
 
After the meshes are built, we assign an index $I \in [1, N_E(n)]$ to each of the $N_E(n)$ triangles $T_{\underset{n}{I}}$ forming the mesh for the $n$-th band. We then define a set of basis functions, where each basis function $\Phi_{\underset{n}{I}}^{i}(\cdot)$ is non-zero only within the $I$-th triangle, and zero everywhere else (the small index $i$ distinguishes basis functions with support over the same triangle). Within a given triangle $I$, we construct all linearly independent orthonormal basis functions that are polynomials up to the linear order. As in $2D$ exactly three basis functions are needed, they are labeled through the small index $i \in [0,2]$ (see~\ref{chapA:basis} and Fig.~\ref{fig:basisfunmed}). Notice that the basis functions $\Phi_{\underset{n}{I}}^{i}(\cdot)$ are discontinuous at the edge of the elements $I$.


Following the approach of spectral methods, we write the distribution functions as a linear combination of basis functions
\begin{equation} \label{eq:fProjdef}
 f_n(t,\bold k) = \sum_{\underset{I}{i}}  f_{\underset{n}{I}}^{i}(t)\; \Phi_{\underset{n}{I}}^{i}(\bold k) \textrm{ ,}
\end{equation}
where $f_{\underset{n}{I}}^{i}(t)$ is the semidiscrete (because the time variable is not discretised yet) representation of the function $f_n(t,\bold k)$.
Thanks to the orthonormality of the basis set and the compact support of each basis function, we have
\begin{equation} \label{eq:fProj}
 f_{\underset{n}{I}}^{i}(t) \equiv \int_{T_{\underset{n}{I}}} \mathrm d^2 \bold k ~ f_n(t,\bold k) \; \Phi_{\underset{n}{I}}^{i}(\bold k) \textrm{ ,}
\end{equation}
where the integration does not need to extend to the full Brillouin zone, but only to the triangle ${T_{\underset{n}{I}}}$ in which the basis function is non-zero.
For later convenience we define
\begin{equation} \label{eq:oneProj}
1_{\underset{n}{I}}^{i} \equiv \int_{T_{\underset{n}{I}}} \mathrm d^2 \bold k ~ \Phi_{\underset{n}{I}}^{i}(\bold k) \textrm{ ,}
\end{equation}
which is the discretised representation of a function that is constant and equal to 1 throughout the domain.

We now project Eq.~\eqref{eq:eecollisionScatNum} onto the basis and apply Eqs.~\eqref{eq:fProjdef}, \eqref{eq:fProj} and \eqref{eq:oneProj} to get

\begin{strip}
\begin{equation}
\begin{split}
& \left ( \frac{\mathrm d  f_{\underset{n_0}{I}}^{i} }{\mathrm d t} \right )_{n_0 + n_1 \leftrightarrow n_2 + n_3 }  \equiv \int_{T_{\underset{n_0}{I}}} \mathrm d^2 \bold k ~\Phi_{\underset{n_0}{I}}^{i}(\bold k_0) \left ( \frac{\partial f_0}{\partial t} \right )_{n_0 + n_1 \leftrightarrow n_2 + n_3 } =  \\
 & \quad \quad \quad =  \sum_{\underset{J,K,M,N}{j,k,m,n}} \left ( \mathbb S _{n_0 + n_1 \leftrightarrow n_2 + n_3 } \right )^{ijkmn}_{\underset{n_0 n_1 n_2 n_3}{IJKMN}} \Big (  ( 1_{\underset{n_0}{J}}^{j} - f_{\underset{n_0}{J}}^{j}   )   ( 1_{\underset{n_1}{K}}^{k} - f_{\underset{n_1}{K}}^{k}   ) f_{\underset{n_2}{M}}^{m}  f_{\underset{n_3}{N}}^{n}  - f_{\underset{n_0}{J}}^{j} f_{\underset{n_1}{K}}^{k}  ( 1_{\underset{n_2}{M}}^{m} - f_{\underset{n_2}{M}}^{m}   ) ( 1_{\underset{n_3}{N}}^{n} - f_{\underset{n_3}{N}}^{n} ) \Big ),
\end{split} \label{eq:kcol1}
\end{equation}
with 
\begin{equation}
\begin{split}
 \left ( \mathbb S _{n_0 + n_1 \leftrightarrow n_2 + n_3 } \right )^{ijkmn}_{\underset{n_0 n_1 n_2 n_3}{IJKMN}} =  \sum_{\bold G}  \int\displaylimits_{T_{\underset{n_0}{J}}} & \int\displaylimits_{T_{\underset{n_1}{K}}} \int\displaylimits_{T_{\underset{n_2}{M}}} \int\displaylimits_{T_{\underset{n_3}{N}}} \mathrm d^2 \bold k_0\;  \mathrm d^2 \bold k_1\;  \mathrm d^2 \bold k_2\; \mathrm d^2 \bold k_3   ~  w^\textrm{e-e}_{0123} \Phi_{\underset{n_0}{I}}^{i}(\bold k_0)  \Phi_{\underset{n_0}{J}}^{j}(\bold k_0) \Phi_{\underset{n_1}{K}}^{k}(\bold k_1) \Phi_{\underset{n_2}{M}}^{m}(\bold k_2) \Phi_{\underset{n_3}{N}}^{n}(\bold k_3) \times \\ 
 &\times \delta(\bold k_0 +\bold k_1 - \bold k_2 - \bold k_3+ \bold G)  ~\delta \left ( \epsilon_{n_0}(\bold k_0) + \epsilon_{n_1}(\bold k_1) - \epsilon_{n_2}(\bold k_2) -\epsilon_{n_3}(\bold k_3) \right ) \textrm{ ,}
\end{split} \label{eq:kscat} 
\end{equation}
\end{strip}
which we call  scattering tensor. Notice how the integral in Eq.~\eqref{eq:kscat} has already undergone a major simplification: thanks to the compact support of the basis functions, the integral does not extend anymore over the full Brillouin zone four times, but only over the cartesian product of four triangles.

The scattering tensors contains all  information about the scattering. Once known we may calculate the collision integral by merely contracting the tensor with the discretized distribution functions (see Eq.~\eqref{eq:kcol1}). 
Assuming that all involved bands have the same number of basis-functions $N_E$ the number of tensor elements would scale $\propto {N_E}^5$ making realistic calculations in $2D$ impossible. However, thanks to the choice of basis functions, it can be shown (\ref{chapA:scaling}) that the tensor is extremely sparse and the number of non-zero tensor elements scales only with $\propto {N_E}^{2.5}$ in $2D$.  

Notice that each scattering process has an associated scattering tensor. Scattering tensors with a different number of involved states can be constructed and have a similar structure.
For brevity, in the following, we will drop the process name when writing the tensor: 

$ \left ( \mathbb S _{n_0 + n_1 \leftrightarrow n_2 + n_3 } \right )^{ijkmn}_{\underset{n_0 n_1 n_2 n_3}{IJKMN}}$ will be written more compactly as  $\mathbb S ^{ijkmn}_{\underset{n_0 n_1 n_2 n_3}{IJKMN}}$.


\subsection{Calculation of the scattering tensor elements}\label{chap:calcSC}

Each element of the scattering tensor needs to be computed by performing the integral in Eq.~\eqref{eq:kscat}. The main difficulty in computing such integral is the presence of the Dirac deltas.

The first step is to discretize the quasiparticle dispersion $\epsilon_{n}(\bold k)$. We use the same discretization as for the distribution function (for an example see Fig.~\ref{fig:modelbands}). Notice that this transforms the dispersion into a piece-wise linear function. Since each basis function is only non-zero for one domain $I$ and within we have a locally linear dispersion, we can  analytically invert all delta-distributions within the scattering tensor integrals. 

The remaining integration on the momentum- and energy-conserving hypersurface is done with standard Monte Carlo integration (for more details see \ref{chapA:detailsScatTens}). However a finite stochastic error is now present in the numerically constructed scattering tensor elements. As described above, this leads to a breaking of the conservation of energy, particles and momentum within the error bars. Due to this problem, a tensor constructed in this way is not usable for time propagation. We solve this problem by enforcing the conservation of these extensive quantities. 

Particle number, total momentum and total energy can be expressed as appropriately weighted integrals of the population of the type (see \ref{chapA:macroscopQuant} for details) 
\begin{equation}
	\Theta_n (t) = \frac{1}{{(2 \pi)}^2} \int_{V_{BZ}} \mathrm d^2 \bold k ~ \theta_n(\bold k) \; f_n(t,\bold k) \textrm{ ,}
\end{equation}
with the single-particle contribution $\theta_n(\bold k)$ ($\theta_n(\bold k)=1$ for particle number, $\theta_n(\bold k) = \epsilon_n(\bold k)$ for energy, $\theta_n(\bold k) = \bold k$ for momentum; see TABLE~\ref{tab:thermquantity}); $\Theta_n(t)$ is the contribution of band $n$ to the extensive density $\Theta$.
Using the orthonormality of the basis functions one can obtain the simple expression 
\begin{equation}
	\Theta_n (t) =\frac{1}{{(2 \pi)}^2} \sum_{\underset{I}{i}}  \theta_{\underset{n}{I}}^i f_{\underset{n}{I}}^i \textrm{ .}
\end{equation}
It can be shown (\ref{chap:cons_sym}) that requiring $\sum_n d\Theta_n (t) /dt =0$ in Eq.~\eqref{eq:kcol1} is equivalent to requiring

\begin{strip}
\begin{equation}
\begin{split}
0=& \sum_i \Bigg [\theta_{\underset{n_0}{J}}^i  ~\left ( \mathbb S ^{ijkmn}_{\underset{n_0 n_1 n_2 n_3}{JJKMN}} + \mathbb S ^{ijknm}_{\underset{n_0 n_1 n_3 n_2}{JJKNM}} \right ) + \theta_{\underset{n_1}{K}}^i  ~\left ( \mathbb S ^{ikjmn}_{\underset{n_1 n_0 n_2 n_3}{KKJMN}} + \mathbb S ^{ikjnm}_{\underset{n_1 n_0 n_3 n_2}{KKJNM}} \right )\\
&- \theta_{\underset{n_2}{M}}^i  ~\left ( \mathbb S ^{imnjk}_{\underset{n_2 n_3 n_0 n_1}{MMNJK}} + \mathbb S ^{imnkj}_{\underset{n_2 n_3 n_1 n_0}{MMNKJ}} \right ) -  \theta_{\underset{n_3}{N}}^i  ~\left ( \mathbb S ^{inmjk}_{\underset{n_3 n_2 n_0 n_1}{NNMJK}} + \mathbb S ^{inmkj}_{\underset{n_3 n_2 n_1 n_0}{NNMKJ}} \right )\Bigg] \textrm{ .}
\end{split} \label{eq:exconsLong2main}
\end{equation}
\end{strip}
The above equations can be written as coefficient vectors multiplied with the vector representation of the involved scattering tensor elements. The equations are exactly fulfilled if the scattering tensor rewritten as a vector is completely orthogonal to the coefficient vectors. We enforce this orthogonality by Gram-Schmitt orthogonalization (for more details see \ref{chapA:cleanup_details}).

\subsection{Time propagation}\label{chap:timeprop}

The expression in Eq.~\eqref{eq:kcol1} is only semidiscrete, as the time variable is still appearing as a continuous dependence. The structure of Eq.~\eqref{eq:kcol1} is that of a system of first order non-linear ordinary differential equations. Many efficient algorithms are available to propagate such system. We use here a fourth order Runge-Kutta method to time propagate the distribution functions.

\subsection{Scattering rates}\label{chap:scrates}

The discretised scattering rates are obtained by projection of Eq.~\eqref{eq:eecollisionScatRate6Analytic} leading to the expression:

\begin{strip}
\begin{equation}
\lambda_{\underset{n_0}{I}}^i =  \sum_{\underset{J,K,M,N}{j,k,m,n}} \mathbb S ^{ijkmn}_{\underset{n_0 n_1 n_2 n_3}{IJKMN}}
  \Big ( 1_{\underset{n_0}{J}}^j  \big (1_{\underset{n_1}{K}}^k - [f_\textrm{eq}]_{\underset{n_1}{K}}^k  \big ) [f_\textrm{eq}]_{\underset{n_2}{M}}^m [f_\textrm{eq}]_{\underset{n_3}{N}}^n + 1_{\underset{n_0}{J}}^j [f_\textrm{eq}]_{\underset{n_1}{K}}^k \big (1_{\underset{n_2}{M}}^m - [f_\textrm{eq}]_{\underset{n_2}{M}}^m \big ) \big (1_{\underset{n_3}{N}}^n - [f_\textrm{eq}]_{\underset{n_3}{N}}^n \big )   \Big ) \textrm{ .}\label{eq:eecollisionScatRate6main}
\end{equation}
\end{strip}
where $f_\textrm{eq}$ is the appropriate equilibrium distribution for that particle type, and $[f_\textrm{eq}]_{\underset{n_1}{K}}^k$ its numerical representation.

These will be used in the following to analyze the scattering channels before doing the actual time propagation.

\section{Non-equilibrium dynamics of model systems}\label{chap:nonequdyn}
In this section we discuss several prototypical non- equilibrium thermalizations and highlight different aspects of the dynamics in order to show the full capabilities of the method.

\begin{figure}
 \centering
 \includegraphics[width=0.4\textwidth]{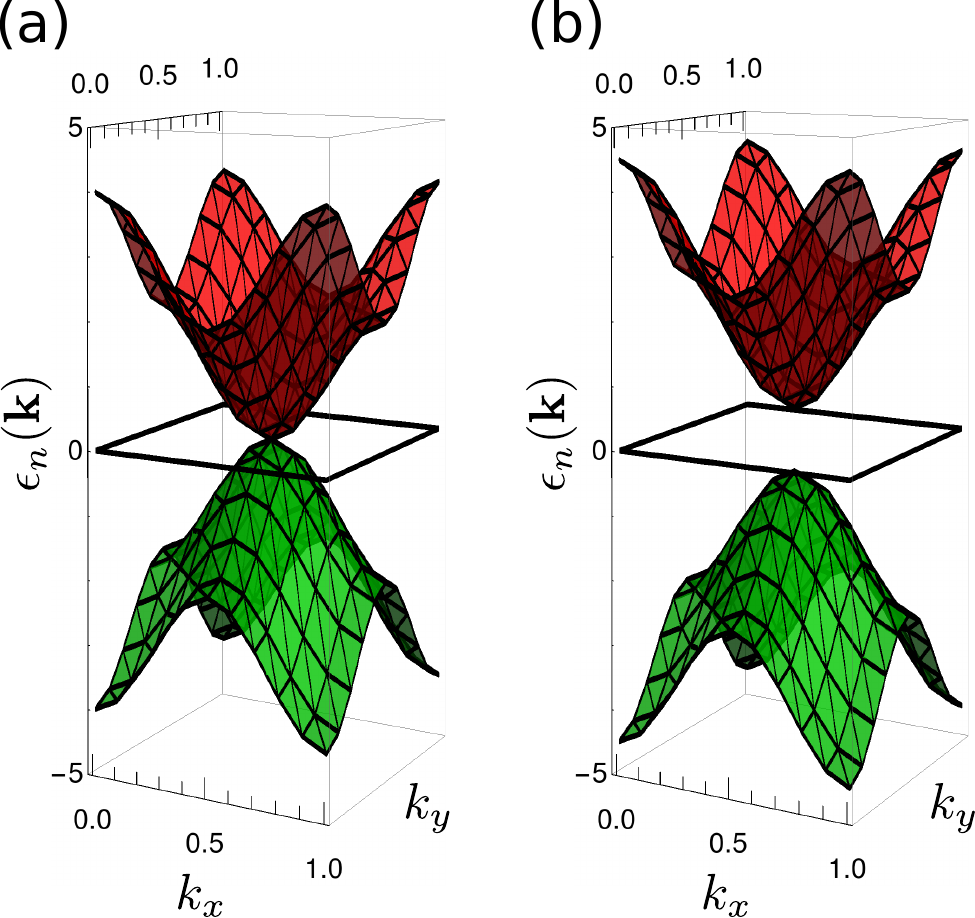}
\caption{Locally linearized band-structure used for thermalization calculations with band-gaps (a) $\Delta = 0$ and (b) $\Delta=1$. For each band we use the same mesh consisting of $N_E=200$ triangles resulting in $N_B=200\times3\times2= 1200$ basis functions in total.}
  \label{fig:modelbands}
\end{figure}

We describe a 2D system, with two electronic bands with the following dispersion relations
\begin{subequations}
\begin{align}
\epsilon_2(\bold k) &\equiv  2 t \cos(2 \pi k_x) + 2 t \cos(2 \pi k_y) + 4 t + \frac{\Delta}{2} \textrm{ ,}\\
\epsilon_1(\bold k) &\equiv -\epsilon_2(\bold k) \textrm{ ,}
\end{align}
\end{subequations}
with the band-gap $\Delta$ and the tight-binding hopping $t = {1}/{2}$ (see Fig.~\ref{fig:modelbands}). Here, we have used a rescaled first Brillouin-zone that occupies the domain $[0,1]\times[0,1]$ instead of $[-\frac{\pi}{2},\frac{\pi}{2}]\times[-\frac{\pi}{2},\frac{\pi}{2}]$. In this notation the $\Gamma$-point is $\bold k_\Gamma = (0.5,0.5)^\textrm{T}$.

\begin{table}[b]
\centering
\resizebox{0.4\textwidth}{!}{
\begin{tabular}{|c|c|}
\hline
 process & description \\
 \hline
 \hline
 $1+ 1 \leftrightarrow 1 + 1$ & scattering within band 1\\ 
 \hline
 $2+ 2 \leftrightarrow 2 + 2$ & scattering within band 2\\ 
 \hline
 $1+ 2 \leftrightarrow 1 + 2$ & scattering between band 1 and 2\\ 
 \hline
 $1+ 1 \leftrightarrow 1 + 2$ &Auger process\\ 
 \hline
 $2+ 2 \leftrightarrow 2 + 1$ &impact excitation\\ 
 \hline
\end{tabular} }
\caption{All possible electron-electron scattering processes for a two-band system; the process $1+1 \leftrightarrow 2+2$ is energetically forbidden.}  \label{tab:eescatteringProc}
\end{table}

\begin{figure*}
 \centering
 \includegraphics[width=0.8\textwidth]{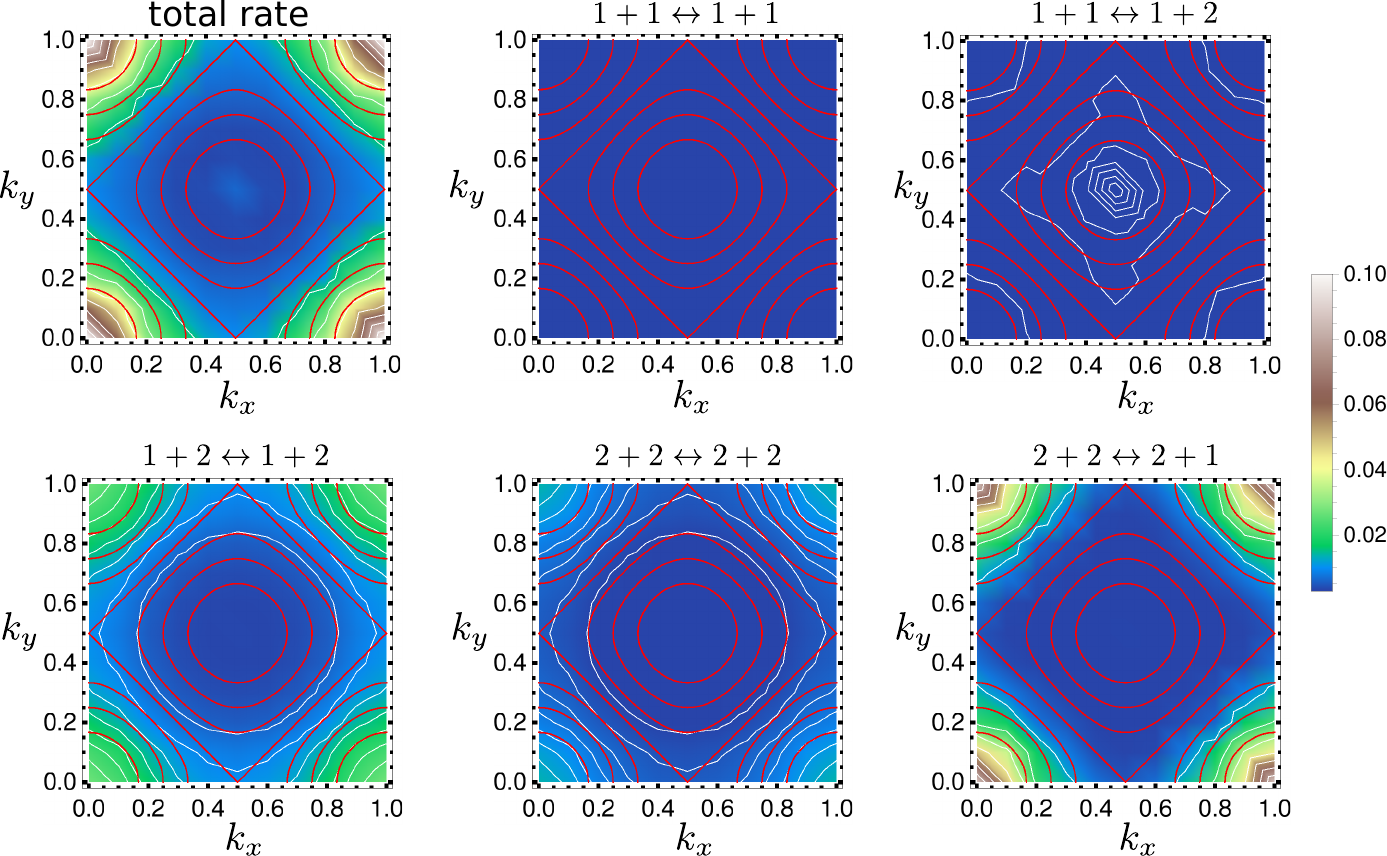}
\caption{Equilibrium scattering rates $\lambda_2(\bold k)$ of an electrons in the upper band (band 2) for the different processes (band gap $\Delta =0$, temperature $T=\frac{1}{6 t} \equiv \frac{1}{3}$). On top the equal-energy lines (red) are plotted as well as the equal-scattering-rate lines (white).}
  \label{fig:200elgap0phsymmSC}
\end{figure*}

We will study the thermalization of an excited system for two different initial strongly out-of-equilibrium distributions and two different band-gaps ($\Delta=0$ and $\Delta = 2 t \equiv 1$).
We include all possible electron-electron scatterings which are shown in TABLE~\ref{tab:eescatteringProc}. Notice that the scattering $1+ 1 \leftrightarrow 2 + 2$ will result in an empty phase space, as there are no transitions that can satisfy energy conservation.

In analogy to the Hubbard model where the interaction is completely local, we assume in the following momentum-independent transition-matrix elements, which is equivalent to momentum-independent scattering amplitudes $w^\textrm{e-e}_{0123}$. Assuming also the same scattering amplitude for all bands, we get a constant scattering amplitude: $w^\textrm{e-e}_{0123} \to w^\textrm{e-e}$.
The value of the scattering amplitude simply determines the global timescale, hence, without loss of generality we choose $w^\textrm{e-e}=1$. Furthermore we do not consider spin in our calculations, i.e. we only have one electron-distribution per band.

For all studied cases, we use Fermi-Dirac distributions with $\mu = 0$ and $\beta = 3$ with additional band resolved excitations $\delta f_n (\bold k)$ as the initial distributions 
\begin{subequations}
\begin{align}
f_2(\bold k,t=0) &= f_\textrm{FD}(\epsilon_2(\bold k), \mu,\beta ) + \delta f_2(\bold k) \textrm{ ,} \\
f_1(\bold k,t=0) &= f_\textrm{FD}(\epsilon_1(\bold k), \mu,\beta ) + \delta f_1(\bold k) \textrm{ .}
\end{align} \label{eq:initiDistr}
\end{subequations}
These may arise e.g. from a laser excitation at momentum $\bold k$ which would lead to $\delta f_2(\bold k) = - \delta f_1(\bold k)$.

\subsection{Scattering rates} \label{chap:scatteringRatesTest}

\begin{figure*}[tb]
 \centering
 \includegraphics[width=0.8\textwidth]{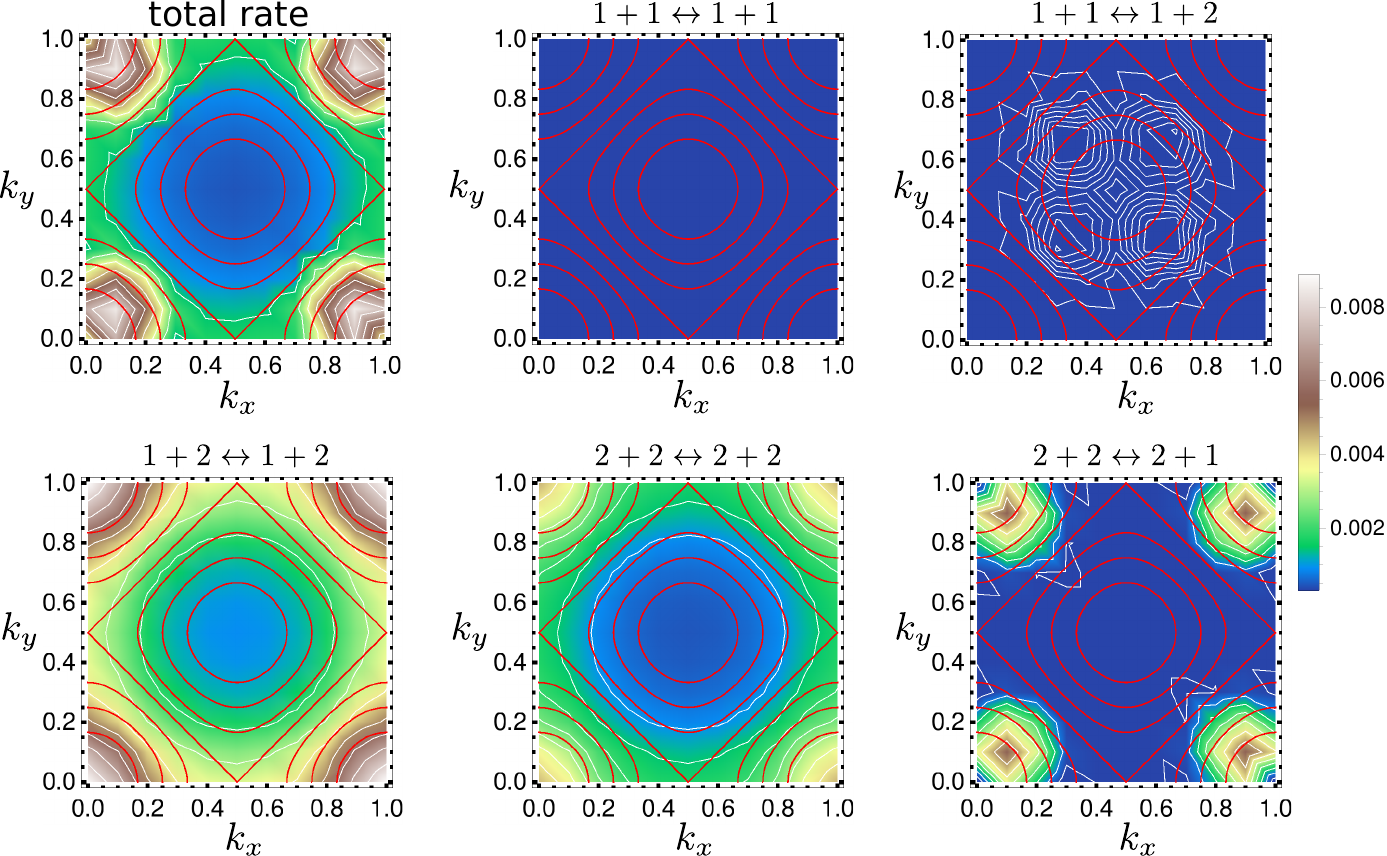}
\caption{Same as Fig.~\ref{fig:200elgap0phsymmSC} but for the band gap $\Delta= 1$. Note the order of magnitude difference in the color scale.}
  \label{fig:200elgap1phsymmSC}
\end{figure*}

Before discussing the full thermalization process, it is instructive to get a preliminary idea of the scattering processes. The dynamics of a scattering process is mainly dictated by the phase space factor, which changes during strongly out-of-equilibrium dynamics. It is in general hard to visualize the internal structure of the scattering tensors which closely resembles the phase space factor, as they are high dimensional functions. Nonetheless often looking at scattering rates close to equilibrium can be an effective driver of intuition even further away from equilibrium.

With Eq.~\eqref{eq:eecollisionScatRate6main} we can calculate the scattering rates of a single electron (or hole) added to the equilibrium system (i.e. $\delta f_1(\cdot) \to 0$; $\delta f_2(\cdot) \to 0$ in Eq.~\eqref{eq:initiDistr}). As the band-structure is particle-hole symmetric, we only discuss the scattering rates of an electron added to the upper band (band 2).

In general, for the gap-less ($\Delta=0$) system, the scattering rates for all the scattering channels become higher with increasing energy (Fig.~\ref{fig:200elgap0phsymmSC}). The level-lines roughly follow the equal-energy lines of the dispersion indicating that the scattering rates mainly depend on the energy of an excitation and not on the momentum explicitly. An exception is the scattering rate of an electron in the upper band due to Auger-process (i.e. the process $1+1 \leftrightarrow 1+2$) which decreases with increasing energy. 
Impact excitation ($2+2 \leftrightarrow 2+1$) is the strongest process, leading to a quick particle tranfer between the bands. Obviously scattering within the lowest band ($1+1 \leftrightarrow 1+1$) does not contribute to the decay of an excitation in band 2, and the associated scattering rates are identically 0 (they have been plotted for completeness and consistency).

The situation is different for the gapped ($\Delta=1$) system (Fig.~\ref{fig:200elgap1phsymmSC}). The scattering rates (except Auger-emission) still increase with increasing energy. The total scattering rate, however, is not anymore approximately only energy dependent (the equal-rate lines in the total rate in Fig.~\ref{fig:200elgap1phsymmSC} do not follow anymore the  equal-energy lines). This indicates a momentum-dependence beyond the dependence through the dispersion-relations. This behavior mainly stems from impact excitation ($2+2 \leftrightarrow 2+1$).

Furthermore, we observe that impact excitation is  now weaker (relative to the other processes) compared to the gap-less system. This is due to the fact that an electron has to be excited across the gap when impact excitation is performed. The larger the gap, the smaller the region within the upper bands where electrons have enough energy (relative to the $\Gamma$-point energy) to excite an electron from the lower band. The allowed phase space for the other relevant processes (i.e. $2+2\leftrightarrow 2+2$ and $1+2 \leftrightarrow 1+2$) is not affected by the gap at all. Yet these two processes are weakened compared to the gap-less system since there are fewer thermally excited electrons (and holes) as scattering partners as the gap $\Delta = 1$ is larger than the fixed temperature $T = \frac{1}{3}$.

Let us stress that the structure of the scattering rates entirely comes from the scattering phase-space and not from transition matrix elements (which we have assumed to be momentum independent).

\subsection{Particle-hole symmetric excitation: $\Delta=0$} \label{chap:partholesymdel0}

\begin{figure*}[tb]
 \centering
 \includegraphics[width=0.8\textwidth]{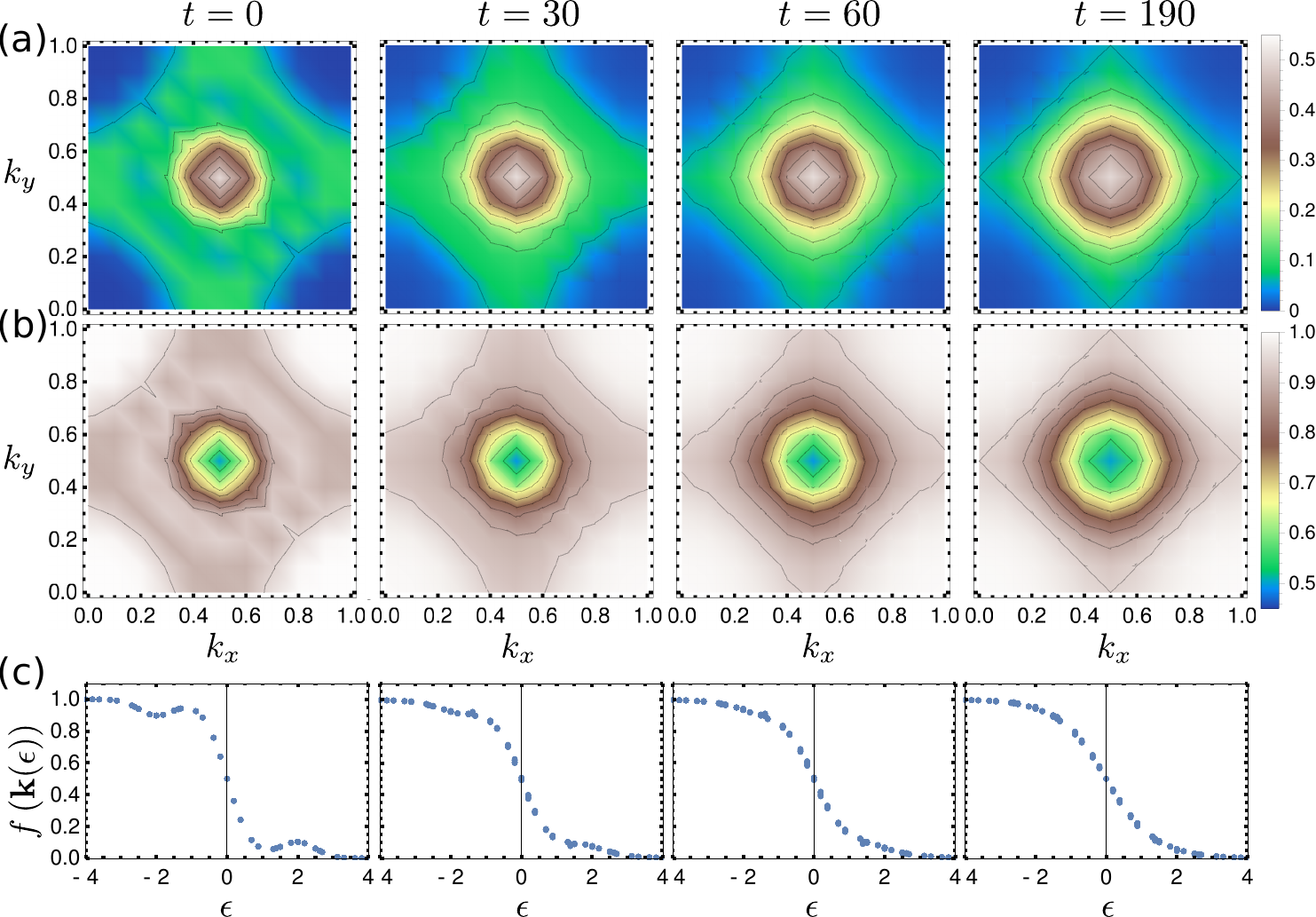}
\caption{Distrubution function $f(\bold k,t)$ of band 2 (a) and band 1 (b) (band gap $\Delta = 0$) for different times. (c) Distribution function as a function of energy for different times. The initial distribution was a Fermi-Dirac distribution with $\mu = 0$, $\beta = 3$ and an excitation (Eq.~\eqref{eq:exc1}) that only depends on the energy.}
  \label{fig:200elgap0phsymm}
\end{figure*}

We now compute the full time propagation, in the presence of the above mentioned scattering channels, of some initial non-equilibrium distributions for the two considered model systems.

First, we study the thermalization of a particle-hole symmetric excitation that depends on the momentum only through the dispersion relations, 
\begin{subequations}
\begin{align}
\delta f_2(\bold k) &= \alpha \times \exp \left ( \frac{-(\epsilon_2(\bold k) - \epsilon_c )^2 }{2 \sigma^2} \right ) \textrm{ ,}\\
\delta f_1(\bold k) &= -\delta f_2(\bold k)
\end{align} \label{eq:exc1}
\end{subequations}
with $\alpha = 0.1$, $\sigma = 0.5$ and $\epsilon_c = 4t + \frac{\Delta}{2}$. This type of excitation is similar to the excitation generated by a laser at an energy $\hbar \omega = 2 \epsilon_c$ that is resonant with the transition between the center van-Hove singularities of the two $2D$ bands.

We calculate the time-propagation for this setup which is shown in Fig.~\ref{fig:200elgap0phsymm} for the system with band gap $\Delta=0$. The band structure and the initial distributions are particle-hole symmetric. As the electron-electron collision operators do not break it, the particle-hole symmetry is maintained at all times. 

During the thermalization process the high-energetic electrons (or holes) of the initial excitations are transferred towards the $\Gamma$-point ($\bold k_{\Gamma} = {(0.5,0.5)}^\textrm{T}$), losing energy in the process. As the total energy is conserved, additional electrons have to be brought up from the lower band to compensate for this energy loss. Eventually, the system thermalizes to a new Fermi-Dirac with a higher temperature than the initial (see  time $t=190$ in Fig.~\ref{fig:200elgap0phsymm}). 

The approach to the equilibrium distribution is more easily recognized when the distribution function is plotted versus energy, i.e. plotting all $f(\bold k(\epsilon),t)$ for a given $\epsilon$ and $t$ (see Fig.~\ref{fig:200elgap0phsymm}c). 
Note, since the dispersion-relations cannot be inverted globally, we get several different distribution-function values for every energy (stemming from different points in the Brillouin-zone).
In principle, far from equilibrium, there is no guarantee that these points will form a curve, as in general the population depends on $\bold k$ only through the energy solely at equilibrium (this is for instance evident in Fig.~\ref{fig:200elgap0Nonphsymm}c below, where at early times the population plotted as a function of energy does not fall on a line, yet after the thermalization has taken place, a Fermi-Dirac distribution is recovered). As we are studying a case where the initial excitation was only dependent on the energy, the population has this characteristic at the initial timestep ($t=0$ case in Fig.~\ref{fig:200elgap0phsymm}c). Interestingly, even though the scattering operators are explicitly momentum dependent, the population preserves this characteristic throughout the whole thermalization process. This is due to the fact that  the scattering rates for this configuration have shown negligible explicit momentum dependence (as shown in section~\ref{chap:scatteringRatesTest} and Fig.~\ref{fig:200elgap0phsymmSC}).

\subsection{Particle-hole symmetric excitation: $\Delta=1$}\label{chap:partholesymdel1}

\begin{figure*}
 \centering
 \includegraphics[width=0.8\textwidth]{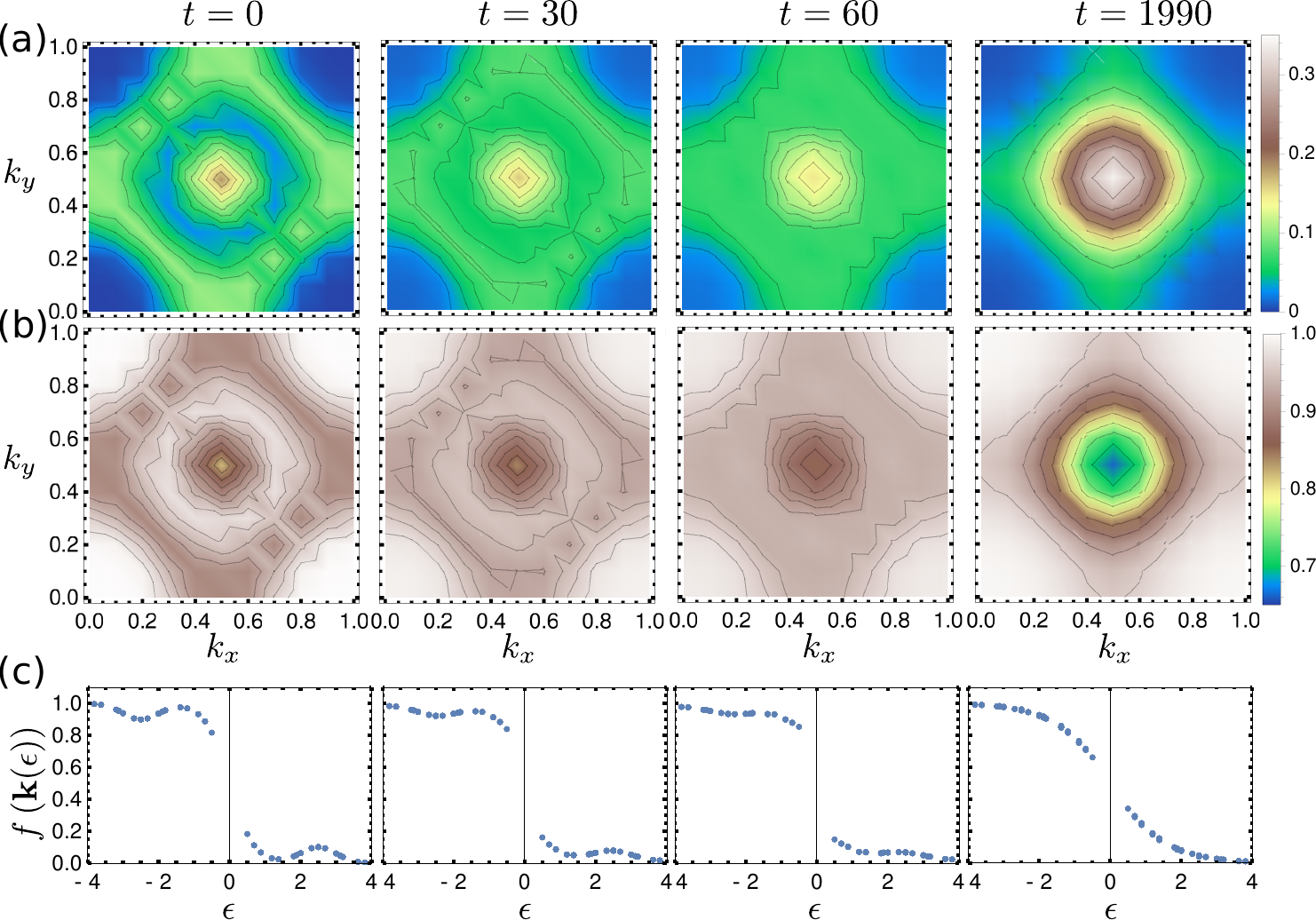}
\caption{Same as Fig.~\ref{fig:200elgap0phsymm} but for a band gap of $\Delta = 1$. Note the much longer thermalization time.}
  \label{fig:200elgap1phsymm}
\end{figure*}

We now address the thermalization dynamics of the system with gap $\Delta = 1$, which shows important qualitative differences compared to the gap-less case. The first difference is that the time needed to thermalize is about one order of magnitude larger than for the system with zero gap. As pointed out in section~\ref{chap:scatteringRatesTest} due to the larger gap there are fewer thermally excited carriers leading to a reduced strength of the processes $1+1 \leftrightarrow 1 +1$, $1+2 \leftrightarrow 1 +2$ and $2+2 \leftrightarrow 2 +2$.
However, the reduced number of thermal carriers alone cannot explain the large difference in the thermalization time. It mainly originates from the fact that the available phase-space for the Auger process and impact excitation is strongly reduced by the  band gap. These are the only processes that may change the number of particles in the bands. In order to reach equilibrium the bands need to transfer particles among each other (which can happen only through the scatterings $1+1 \leftrightarrow 1 +2$ and $2+2 \leftrightarrow 2 +1$). Hence, a complete thermalization can happen only over the time-scales of impact excitation and Auger processes.

This leads yet to another very important effect: the thermalization happens in two distinct steps. The scatterings within the bands ($1+1 \leftrightarrow 1+1$, $2+2 \leftrightarrow 2+2$) and between the bands ($1+2 \leftrightarrow 1+2$) are faster than the remaining impact excitation ($2+2 \leftrightarrow 2 +1$) and Auger process($1+1 \leftrightarrow 1 +2$). As a result the two bands will first undergo an initial partial thermalization, during which they can redistribute energy within each band individually but there is not yet a sizeable number of particles exchanged between the bands. In oder words, the two bands act as two thermodynamic objects that can transfer energy but not particles.

\begin{figure*}
 \centering
 \includegraphics[width=0.8\textwidth]{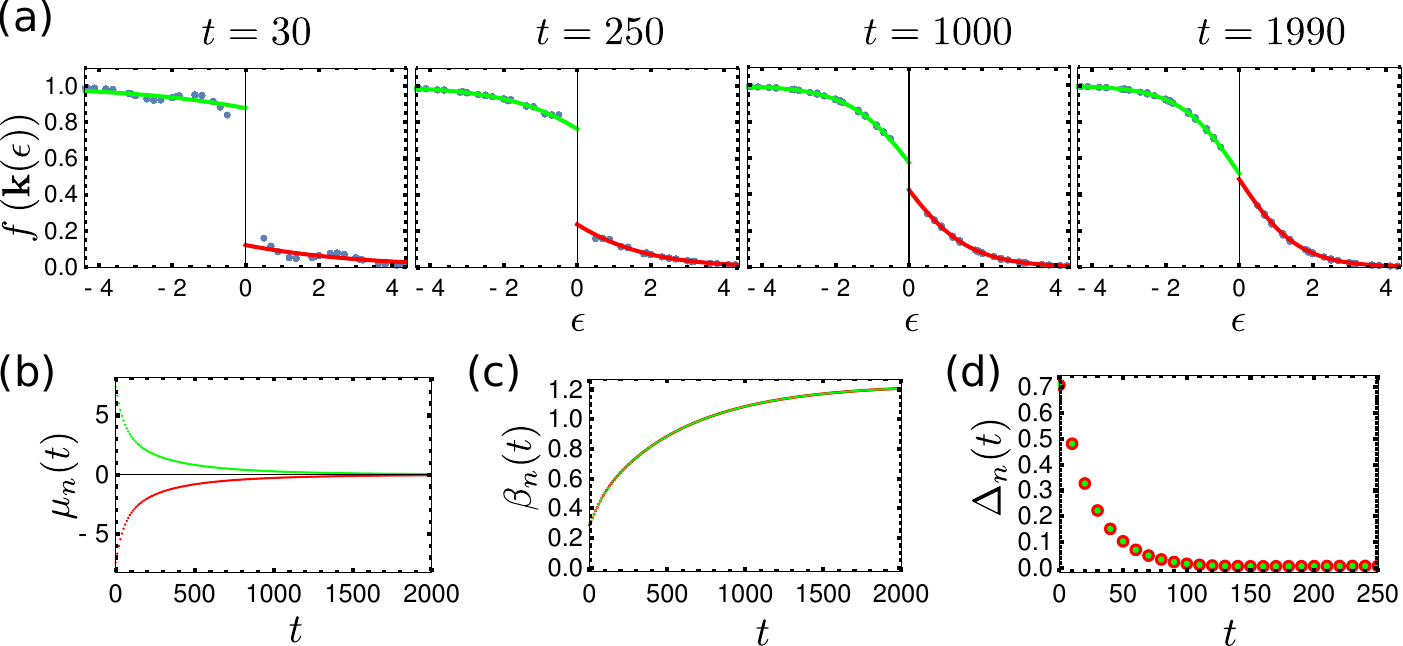}
\caption{(a) Non-equilibrium distribution function $f(\bold k (\epsilon))$ as a function of time $t$ with a Fermi-Dirac fit for the upper- (red) and lower- (green) band. (b) Chemical potentials, (c) inverse temperatures and (d) squared deviation of the Fermi-Dirac fits in dependence of time.}
  \label{fig:200elgap1phsymmFermi}
\end{figure*}

The distribution-functions of the two bands will therefore form two individual Fermi-Dirac distributions with different chemical potentials but the same temperature for times $t \gtrsim 100$. This is visualized in Fig.~\ref{fig:200elgap1phsymmFermi}. In Fig.~\ref{fig:200elgap1phsymmFermi}a we plot the fitting with Fermi-Dirac distributions of the energy resolved population separately for the two bands. At earlier times the fitting error (Fig.~\ref{fig:200elgap1phsymmFermi}d) is too large, showing that the distribution is still far from equilibrium. However, within several tens of time units the two bands already look internally thermalized (as the fitting error drastically decreases in Fig.~\ref{fig:200elgap1phsymmFermi}d).  Within this time the two bands also reach the same temperature because the bands can exchange energy through the process $1+2 \leftrightarrow 1+2$ which is not affected by the band gap (notice that this process is shadowed in this case by the fact that since the excitation is particle-hole symmetric, the population remains particle-hole symmetric throughout the whole dynamics, making the temperature trivially identical). 

Nonetheless one can clearly see how a global thermalization has not been reached yet within the first several hundreds of time units. The two individual Fermi-Diracs have chemical potentials that lie below (above) zero for the upper (lower) band. As time progresses, the two chemical potentials, however, approach each other, due to Auger process and impact excitation scatterings on a scale of 2000 time units. One can also observe how the temperature of the two bands decreases. Eventually both chemical potentials equalize and the system reaches global thermal equilibrium where it can be described with a single Fermi-Dirac distribution for both bands (see time  $t=1990$ in Figs.~\ref{fig:200elgap1phsymm} and \ref{fig:200elgap1phsymmFermi}).

\subsection{Particle-hole asymmetric excitation: $\Delta=0$}\label{chap:partholenonsymdel0}

\begin{figure*}
 \centering
 \includegraphics[width=0.8\textwidth]{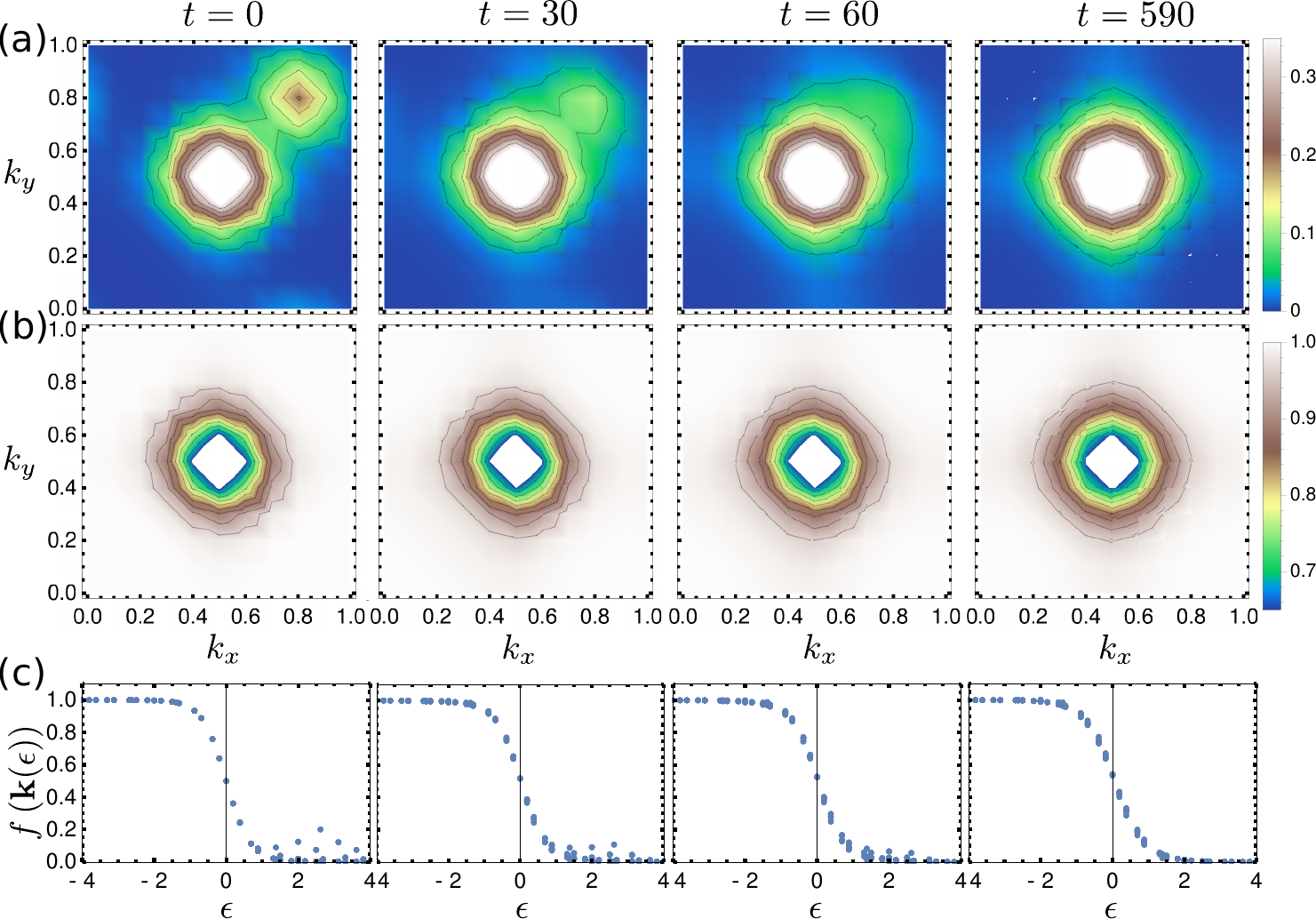}
\caption{Distrubution function $f(\bold k,t)$ of band 2 (a) and band 1 (b) ($\Delta = 0$) for different times. (c) The distribution function in dependence of the energy for different times. The initial distribution was a Fermi-Dirac with $\mu = 0$, $\beta = 3$ and an excitation that explicitly depends on the momentum.}
  \label{fig:200elgap0Nonphsymm}
\end{figure*}

In this section we will discuss a case where the excitation breaks particle-hole symmetry and is momentum asymmetric. We use the same band-structures as in the previous sections (which are particle-hole symmetric) but a different excitation (which is now not particle-hole symmetric and also $\bold k $-dependent),
\begin{subequations}
\begin{align}
\delta f_2(\bold k) &= \alpha  \sum_{\bold G} \exp \left ( \frac{-(\bold k - \bold k_c -\bold G)^2 }{2 \sigma^2} \right )  \textrm{ ,}\\
\delta f_1(\bold k) &= 0 \textrm{ ,}
\end{align} \label{eq:initialDistgap1Nonsym}
\end{subequations}
with $\alpha = 0.2$, $\sigma = 0.1$ and $\bold k_c = {(0.8,0.8)}$. The sum over all reciprocal lattice vectors is needed to ensure that the distribution function is periodic at the borders of the first Brillouin-zone.

\begin{figure}[tb]
 \centering
 \includegraphics[width=0.47\textwidth]{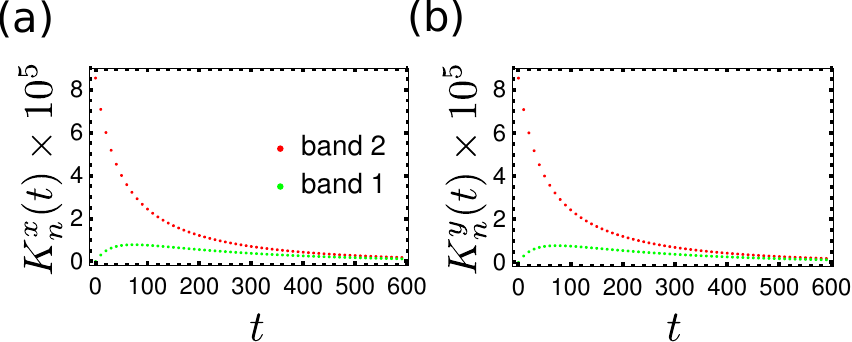}
\caption{Total momentum density (red: upper band; green: lower band) in dependence of time for the $x$- (a) and $y$- (b) momentum component (gap $\Delta = 0$). The initial distribution is symmetric along the $k_x=k_y$ diagonal, hence, the momentum density is equal in $x$- and $y$-direction.}
  \label{fig:200elgap0NonphsymmK}
\end{figure}

First, we discuss the case with zero band gap (Fig.~\ref{fig:200elgap0Nonphsymm}). 
As seen in the previous sections, the electrons are redistributed by the scatterings towards the $\Gamma$-point during the thermalization process and electrons are excited from the lower band to the upper band to compensate for the energy loss. In contrast to the case discussed in the previous sections the distribution function in dependence of the energy (Fig.~\ref{fig:200elgap0Nonphsymm}c) now clearly shows that the population is not a function of energy only. This is a consequence of starting with a distribution function centered around $\bold k_c = {(0.8,0.8)}^\textrm{T}$ which even has a net lattice momentum. To achieve a full thermalization towards a Fermi-Dirac (which has no net momentum), momentum needs to be dissipated. Since no electron-phonon scatterings have been included, this can happen only through two different electronic processes. (i) The electrons in the upper band scatter with each other ($2+2 \leftrightarrow 2+2$) and perform umklapp processes reducing the total momentum. (ii) The electrons of the upper band scatter with electrons of the lower band ($1+2 \leftrightarrow 1+2$) where they transfer momentum from the upper to the lower band and/or dissipate momentum through umklapp processes. The electrons of the lower band will then as well scatter with each other ($1+1 \leftrightarrow 1+1$) and dissipate momentum through further umklapp processes (Fig.~\ref{fig:200elgap0NonphsymmK}). 

We calculate the total momentum density by integrating over the population (see also~\ref{chapA:macroscopQuant})
\begin{equation}
\bold K_n(t) = \int_{V_{BZ}}  \mathrm d^2 k \frac{1}{{(2 \pi)}^2} f_{n}(t, \bold k) \left ( \bold k - {(0.5,0.5)} \right ) \textrm{ ,}
\end{equation}
and plot the results in Fig.~\ref{fig:200elgap0NonphsymmK}. We can see how the momentum of the upper band is partially reduced by umklapp and partially transferred to the lower band. The total momentum in the lower band shows an initial increase due to the direct transfer from the other band. Eventually both tend to decay to a situation with a vanishing total momentum.  Interestingly, one can observe how the dissipation slows down considerably with time. The reason is that when the population decays closer to the $\Gamma$-point, fewer and fewer electrons are still close enough to the edge of the Brillouin zone to perform umklapp.

\begin{figure*}[tb]
 \centering
 \includegraphics[width=1\textwidth]{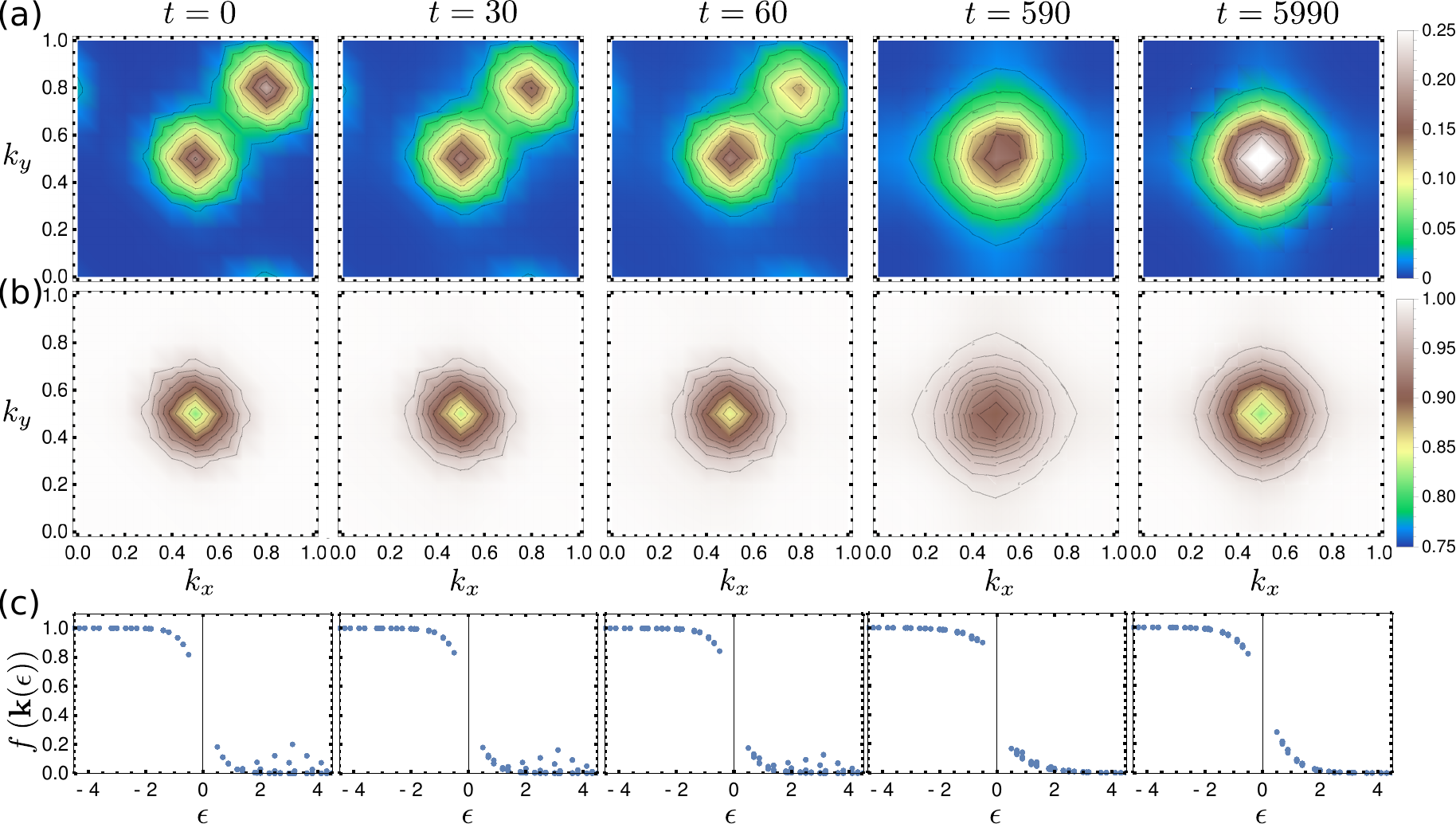}
\caption{Same as Fig.~\ref{fig:200elgap0Nonphsymm} but for a band gap of $\Delta = 1$. Note the much longer thermalization time.}
  \label{fig:200elgap1Nonphsymm}
\end{figure*}

\subsection{Particle-hole asymmetric excitation: $\Delta=1$}\label{chap:partholenonsymdel1}

\begin{figure*}
 \centering
 \includegraphics[width=1\textwidth]{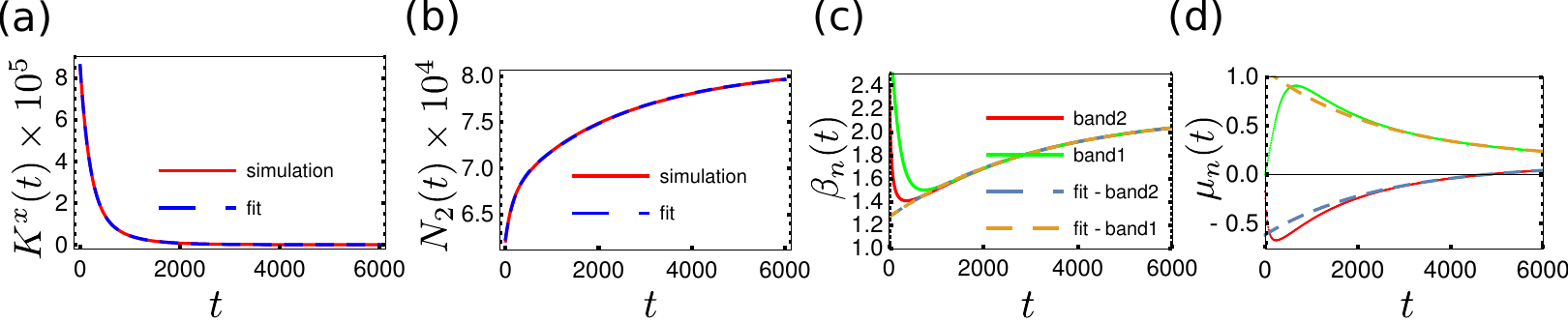}
\caption{(a) Total momentum density in $x$-direction (red) with double-exponential fit (blue) on top, (b) particle density of the upper band (red) with double-exponential fit (blue) on top, (c) inverse temperatures and chemical potentials (d) of Fermi-Dirac fits to the upper (red) and lower (green) band with single-exponential fits (dashed, fitting-interval $t\in [4000,5990]$) on top. }
  \label{fig:200elgap1NonphsymmK}
\end{figure*}

We now simulate the thermalization process of the gapped system ($\Delta = 1$; Fig.~\ref{fig:200elgap1Nonphsymm}) with the same initial distribution (Eq.~\eqref{eq:initialDistgap1Nonsym}). As in the gap-less system, the initial momentum has to be dissipated in order to thermalize the system. In principle, the processes $2+2 \leftrightarrow 2+2$, $1+2 \leftrightarrow 1+2$ and $1+1 \leftrightarrow 1+1$ are not affected by the gap, hence, one might expect that the momentum exchange between the upper and lower band and the dissipation should be as fast as in the gap-less system. However, this is not the case (Fig.~\ref{fig:200elgap1NonphsymmK}a). 
The reason is again the smaller number of thermally excited carriers. Since they involve two electrons, the strengths of the scattering processes relevant for momentum dissipation depend on the number of carriers in the band, both the ones that are thermally present and the excited ones.

The time-dependence of the total momentum in $x$-direction (i.e. $K^x(t)=K^x_1(t) + K^x_2(t)$) can be well described with a double-exponential function ($g(t) = a + b \times \textrm{Exp}(-\frac{t}{\tau^I})) + c \times \textrm{Exp}(-\frac{t}{\tau^{II}})$; with $\tau^{I} \leq \tau^{II}$)(Fig.~\ref{fig:200elgap1NonphsymmK}a). The two times obtained from the fit are $\tau_K^I = 194$ and $\tau_K^{II} = 597 $. We attribute the two different timescales to the strong energy dependence of the scattering rates. The short time $\tau_K^I$ reflects the momentum dissipation of the initial high-energy carriers while the larger time $\tau_K^{II}$ is the average dissipation time of the low energetic electrons closer to the $\Gamma$-point (note that only Umklapp processes contribute to the momentum dissipation).

After a time $t = 3\times \tau_K^{II} \approx 1800$ a large part of the momentum has already decayed and the purely energy dependent representation is justified again. Therefore, for times $t \gtrsim 3\times \tau_K^{II}$ it makes sense to perform Fermi-Dirac fits within each band. As we can see from the time-dependent chemical potentials $\mu_n(t)$ and inverse temperatures $\beta_n(t)$, the system undergoes the same step of partial thermalization as in the previous section where the upper and lower bands are populated according to Fermi-Dirac distributions with the same temperature but different chemical potentials. With increasing time, the chemical potentials approach each other until they equalize and the system reaches global equilibrium. We can estimate the time it takes for global thermalization from single exponential fits (i.e. with $y(t) = a + b \times \textrm{Exp}(-\frac{t}{\tau})$) to $\mu_n(t)$ and $\beta_n(t)$ within the time interval $t \in [4000,5990]$. We get $\tau_{\mu_1} = 2564$, $\tau_{\mu_2} = 2554$, $\tau_{\beta_1} = 3207$ and $\tau_{\beta_2} = 3182$. The two different times we get from the two bands for each quantity are identical within the tolerance; $\tau_{\beta_1} \approx \tau_{\beta_2} \equiv \tau_{\beta}$, $\tau_{\mu_1} \approx \tau_{\mu_2} \equiv \tau_{\mu}$. As expected, chemical potential and inverse temperature thermalize on similar timescales albeit not identical. 

It is interesting to study the particle density in dependence of time. In analogy to the momentum density, the band resolved particle density can be calculated with the relation  
\begin{equation}
N_n(t) = \int_{V_{BZ}}  \mathrm d^2 k \frac{1}{{(2 \pi)}^2} f_{n}(t, \bold k) \textrm{ .}
\end{equation}

From the scattering rates Fig.~\ref{fig:200elgap1phsymmSC} we see, that some of the initial electrons may directly perform impact excitation (i.e. the process $2+2\leftrightarrow 2+1$). These initial, high energy electrons rapidly change the particle number in the upper band (Fig.~\ref{fig:200elgap1NonphsymmK}b). After the initial electrons have decayed to lower energies, the low energetic electrons must perform several scatterings to gain again enough energy for the process $2+2\leftrightarrow 2+1$ leading to a much longer timescale for total thermalization. 
Similar to the momentum density, the particle density follows a double exponential function. From a fit over the whole timescale we get $\tau^I_{N_2}=178$ and $\tau^{II}_{N_2}=2479$.
The scattering rate of the process $2+2\leftrightarrow 2+1$ is around $\lambda_\textrm{impact} = 0.0055$ at its maximum, leading to a lifetime of $\tau_\textrm{impact}= 1/\lambda_\textrm{impact} = 182$ which is approximately $\tau^I_{N_2}$. The larger thermalization time $\tau^{II}_{N_2}$ reflects the long-time thermalization of the system and is approximately the same as the time constant $\tau_{\mu}$ determined from the fitting of the time-dependence of chemical potentials.

Summarizing, the case of the gapped system with explicitly momentum dependent initial distribution reveals dynamics on several timescales. First, the initial, finite momentum is dissipated, i.e.~the electrons that were added in a finite region in momentum space are distributed symmetrically over the whole Brillouin-zone. This process takes place on two timescales, one for high energetic electrons ($\tau_K^I$) and one for long-time dissipation ($\tau_K^{II}$). The high energetic electrons of the upper band also perform impact excitation which quickly increases the number of particles in the upper band ($\tau_{N_2}^I$). 
Then the upper and lower band behave like two separately thermalized systems with different chemical potentials. Only high-energetic electrons or holes, which are few at that time, may perform processes that lead to a particle transfer between the sub-systems. Moreover, after a high energetic electron has brought up an electron through impact excitation, both electrons end up with low energy. They need to undergo further several scatterings to get enough kinetic energy to perform another impact excitation. This determines the timescale on which the chemical potentials of the upper and lower band equilibrate ($\tau_{\mu} \approx \tau_{N_2}^{II}$), reaching full thermalization. The timescales can be set in relation to each other  giving $\tau_{N_2}^I < \tau_K^I < \tau_K^{II} < \tau_{\mu}$.




\section{Conclusion}\label{chap:conclusion}

We have introduced an innovative numerical  method to solve the most challenging part of the time-dependent Boltzmann equation, the scattering term, without any approximation in the strongly-out-of-equilibrium regime. We have achieved a vastly improved scaling of the computational cost with respect to precision compared to straightforward implementations. The method conserves to machine precision particle number, momentum and energy at any resolution, allowing for time propagation till full thermalization. The method allows for the use of realistic band structures, multiple types of quasiparticles and multiple types of scattering channels. Finally it can be straightforwardly complemented with a deterministic solver for the transport.

We have applied the numerical method to solve the thermalization dynamics under different types of scatterings in two cases: a metal and a semiconductor. We computed the scattering rates and scattering lifetimes for all the scattering channels analyzing their momentum and energy dependence. We analyzed how the system evolves from a strongly out-of-equilibrium perturbation of an initial equilibrium towards a Fermi-Dirac at increased temperature and possibly altered chemical potential. We also showed how different timescales arise in the semiconducting case and the system first achieves a partial thermalization before fully thermalizing.

To our knowledge no other numerical techniques have been so far able to produce a solution to the  time-dependent Boltzmann equation, to this level of complexity, free from close-to-equilibrium approximations, for realistic band structures and, especially, for this high order scattering channels. This approach will have a major impact in the description of ultrafast dynamics in solids, by making a full ab initio description of the full dynamics finally possible.


\section*{Acknowledgments}
M.W.~and M.B.~acknowledge Nanyang Technological University, NAP-SUG; M.W. acknowledges FWF for funding through Doctoral School W1243 Solids4Fun (Building Solids for Function); and K.H. the FWF for support through project P30997.

\appendix
\section{Definition of the basis functions} \label{chapA:basis}
The definition of the $2D$ basis-functions as described in the main text reads
\begin{equation} 
 \Phi_{\underset{n}{I}}^{i}(\bold k)= \left\{
\begin{array}{ll}
     \mathcal P_{\underset{n}{I}}^i ( \bold k) & \quad \bold k \in T_{\underset{n}{I}} \\ \label{eq:dgBasis2D}
    0 & \quad \textrm{otherwise}
\end{array} \right. ,
\end{equation}  
with 
\begin{equation}
 \mathcal P_{\underset{n}{I}}^i ( \bold k)= \left\{
\begin{array}{ll}
      \gamma^0_{\underset{n}{I}}  & \quad i=0 \\ \label{eq:dgBasis2D_2}
     \beta^1_{\underset{n}{I}} k_y +  \gamma^1_{\underset{n}{I}} & \quad i=1\\
     \alpha^2_{\underset{n}{I}} k_x + \beta^2_{\underset{n}{I}} k_y +  \gamma^2_{\underset{n}{I}} & \quad i=2
\end{array} \right. .   
\end{equation}  
The above definition contains six unknown coefficients per element that are determined by  requiring  orthonormality, i.e.
\begin{equation}
 \int \mathrm d ^2 k ~ \Phi_{\underset{n}{I}}^{i}(\bold k) \Phi_{\underset{n}{J}}^{j}(\bold k) = \delta_{I,J} \delta_{i,j} \textrm{ .}
\end{equation}
A basis of this type is commonly used in the so-called Discontinuous Galerkin (DG) finitie-elements methods which is the reason why we will call it DG-basis in the following.

\section{Calculation of macroscopic quantities}\label{chapA:macroscopQuant}
Within the Boltzmann framework the band-resolved contribution to an extensive thermodynamic density $\Theta_{n}$ can be calculated with
\begin{equation}
\Theta_{n}(t) =  \int_{V_{BZ}}  \mathrm d^d k \frac{1}{{(2 \pi)}^d} f_{n}(t, \bold x, \bold k) \theta_{n}(\bold x, \bold k) \textrm{ .}\label{eq:thermquantity}
\end{equation}
where $\theta_{n}(\bold x, \bold k)$ is the single-particle contribution (see TABLE~\ref{tab:thermquantity}).
In the basis introduced above Eq.~\eqref{eq:thermquantity} becomes 
\begin{equation}
\Theta_n = \frac{1}{{(2 \pi)}^d}\sum_{I,i} \theta_{\underset{n}{I}}^i a_{\underset{n}{I}}^i \textrm{ .}\label{eq:singlePartProj}
\end{equation}
where $\theta_{\underset{n}{I}}^i$ is the expansion coefficient of the single-particle contribution in the basis.

\begin{table}[h]
\centering

\begin{tabular}{|l|c|r|}
\hline
 description & $\Theta$ & $\theta_{n}(\bold x, \bold k)$ \\
 \hline
 \hline
 particle density  & $N$ & 1\\ 
 \hline
 charge density & $C$ & $-e$\\ 
 \hline
 spin density & $\mathcal S$ & $\sigma$ \\ 
 \hline
  momentum density & $\bold K$ & $\bold k - \bold k_{\Gamma}$ \\ 
 \hline
 inner energy density & $E$ & $\epsilon_{n}(\bold k)$\\ 
 \hline
 total energy density& $U$ & $\epsilon_{n}(\bold k) - e \phi(\bold x)$\\ 
 \hline
\color{gray} entropy density &\color{gray} $S$ &\color{gray} $\left ( \epsilon_{n}(\bold k) - \mu(\bold x) \right ) / T(\bold x)$\\ 
\hline
\color{gray} heat density &\color{gray} $Q$ &\color{gray} $ \epsilon_{n}(\bold k) - \mu(\bold x) $\\ 
 \hline
\end{tabular} 
\caption{Table of different extensive, thermodynamic densities, their associated symbols $\Theta$ and their corresponding single-particle contribution $\theta_{n}(\bold x, \bold k)$ for Eqs. \eqref{eq:thermquantity}. $\phi(\bold x)$ is the local electrical potential, $T(\bold x)$ is the local temperature and $\mu(\bold x)$ is the local chemical potential.}  \label{tab:thermquantity}
\end{table}
 
\section{Scaling of the scattering tensor} \label{chapA:scaling}
In general, each band can have a different number of basis functions. In order to understand the scaling of the scattering tensor we will assume here that all bands involved in the scattering have the same amount of basis functions that is proportional to the number of mesh elements $N_E$. At first glance the number of tensor elements scales with the number of mesh elements as $N_{SC} \propto {N_E}^5$. This is indeed the case when an arbitrary basis is used.
However, as the DG-basis has the property that every basis function has a compact support the effective scaling is strongly improved. The reason is the following: The integration domain in the scattering tensor is $4d$-dimensional for a system of spatial dimension $d$. However, the integral contains $d+1$ delta-distributions limiting the effective integration to the $4d - (d+1) = 3d-1$ dimensional hypersurface that conserves the energy and the momentum. Each dimension is covered by ${N_E}^{\frac{1}{d}}$ basis functions (which is only true due to their compact support), hence the effective scaling (i.e. only non-zero elements) of the four-leg scattering tensor is 
\begin{equation}
N_{SC}^\textrm{4-leg} \propto {\left ( {N_E}^{\frac{1}{d}} \right ) }^{3d - 1} = {N_E}^{3-\frac{1}{d}} \textrm{ .}
\end{equation} 

\section{More details on the numerical calculation of the tensor elements}\label{chapA:detailsScatTens}
As each DG-basis function is only non-zero within one element, the actual integration domain for four momenta is not ${V_{BZ}} \times {V_{BZ}} \times {V_{BZ}} \times {V_{BZ}} $ but rather $T_{\underset{n_0}{J}} \times T_{\underset{n_1}{K}} \times T_{\underset{n_2}{M}} \times T_{\underset{n_3}{N}}$, i.e. the cartesian product of the corresponding triangles (in general elements). As all the momenta inside the integral are each one limited to a single triangle, we can approximate the corresponding dispersion relations with their linearized versions. For instance for the first state involved in the scattering we can write
\begin{equation}\label{eq:epslin}
\epsilon_{n_0}(\bold k_0) |_{\bold k_0 \in  T_{\underset{n_0}{J}}} \to \bar \epsilon_{\underset{n_0}{J}}(\bold k_0) \equiv \bold u_{\underset{n_0}{J}} \cdot \bold k_0 + t_{\underset{n_0}{J}} \textrm{ ,}
\end{equation}
and equivalently for the other states. The three coefficients $\bold u_{\underset{n_0}{J}}$ and $t_{\underset{n_0}{J}}$ are exactly determined by the requirement that the linearized dispersion $\bar \epsilon_{\underset{n_0}{J}}(\bold k_0)$ is equal to the original dispersion $\epsilon_{n_0}(\bold k_0)$ at the three nodes of the triangle $T_{\underset{n_0}{J}}$ (this leaves the locally linearized dispersions a globally continuous function, as in  Fig.~\ref{fig:modelbands}). 


Now that the energy-conserving delta-distribution only contains a function that depends linearly on the momenta (that are the integration variables), we can analytically invert it. As a first step we invert the momentum-conserving delta in Eq.~\eqref{eq:kscat} with respect to $\bold k_1$ without loss of generality, which reduces Eq.~\eqref{eq:kscat} to

\begin{strip}
\begin{equation}
\begin{split}
 &\left ( \mathbb S _{n_0 + n_1 \leftrightarrow n_2 + n_3 } \right )^{ijkmn}_{\underset{n_0 n_1 n_2 n_3}{IJKMN}} =  \delta_{I,J} \sum_{\bold G}  \int_{T_{\underset{n_0}{J}}}  \mathrm d^2 k_0   \int_{T_{\underset{n_2}{M}}}  \mathrm d^2 k_2 \int_{T_{\underset{n_3}{N}}} \mathrm d^2 k_3 ~w_{0123}^\textrm{e-e} ~\Omega( (\bold k_2 + \bold k_3 - \bold k_0 - \bold G) \in T_{\underset{n_1}{K}} ) \\
 & \quad    \times  \mathcal P_{\underset{n_0}{I}}^{i}(\bold k_0)  \mathcal P_{\underset{n_0}{J}}^{j}(\bold k_0)  \mathcal P_{\underset{n_1}{K}}^{k}(\bold k_2 + \bold k_3 - \bold k_0 - \bold G)  \mathcal P_{\underset{n_2}{M}}^{m}(\bold k_2)  \mathcal P_{\underset{n_3}{N}}^{n}(\bold k_3)  
  \delta \left ( \bar \epsilon_{\underset{n_0}{J}}(\bold k_0) + \bar \epsilon_{\underset{n_1}{K}}(\bold k_2 + \bold k_3 - \bold k_0 - \bold G) - \bar \epsilon_{\underset{n_2}{M}}(\bold k_2) - \bar \epsilon_{\underset{n_3}{N}}(\bold k_3) \right ) \textrm{,}
\end{split} \label{eq:kscat2} 
\end{equation}
\end{strip}
with the function $\Omega(\cdot)$ that we define to be 1 if the statement inside is true and 0 otherwise. In Eq.~\eqref{eq:kscat2} we have also replaced the basis functions $\Phi$ with the polynomials $\mathcal P$ (see Eq.~\eqref{eq:dgBasis2D}) and we have already restricted the integration domains to the corresponding triangles. 

Using the definition of the linerized dispersion Eq.~\eqref{eq:epslin} the energy-conserving delta in Eq.\eqref{eq:kscat2} reads,

\begin{strip}
\begin{equation} \label{eq:enconslin}
\delta \left (\left ( \bold u_{\underset{n_0}{J}} - \bold u_{\underset{n_1}{K}}  \right ) \cdot \bold k_0 + \left ( \bold u_{\underset{n_1}{K}} - \bold u_{\underset{n_2}{M}}  \right ) \cdot \bold k_2  + \left ( \bold u_{\underset{n_1}{K}} - \bold u_{\underset{n_3}{N}}  \right ) \cdot \bold k_3 + t_{\underset{n_0}{J}} + t_{\underset{n_1}{K}} - t_{\underset{n_2}{M}} - t_{\underset{n_3}{N}} - \bold u_{\underset{n_1}{K}} \cdot \bold G \right ) \textrm{.}
\end{equation}
\end{strip}

For the following let us assume that $\left (  u_{\underset{n_1}{K}}^x -  u_{\underset{n_2}{M}}^x  \right ) \neq 0$ (one can also use any other vector-component of the momentum-prefactors occurring in Eq.~\eqref{eq:enconslin} as long as it is non-zero). When this requirement is fulfilled we can invert the energy-delta with respect to $k_2^x$, which gives

\begin{strip}
\begin{equation}
\begin{split}
 &\left ( \mathbb S _{n_0 + n_1 \leftrightarrow n_2 + n_3 } \right )^{ijkmn}_{\underset{n_0 n_1 n_2 n_3}{IJKMN}} =  \delta_{I,J} \sum_{\bold G}  \int_{T_{\underset{n_0}{J}}}  \mathrm d^2 k_0   \int_{L_{\underset{n_2}{M}}^{[a,b]}}  \mathrm d k_2^y \int_{T_{\underset{n_3}{N}}} \mathrm d^2 k_3 \frac{1}{\big | u_{\underset{n_1}{K}}^x -  u_{\underset{n_2}{M}}^x \big |}  w_{0123}^\textrm{e-e} ~\Omega( (\bold k_2 + \bold k_3 - \bold k_0 - \bold G) \in T_{\underset{n_1}{K}} )|_{\underset{k_2^x = \xi}{}}  \\ 
 &\quad \quad\quad \quad \quad \quad \quad \quad \quad \quad \times \Omega(\bold k_2 \in T_{\underset{n_2}{M}} )|_{\underset{k_2^x = \xi}{}} \mathcal P_{\underset{n_0}{I}}^{i}(\bold k_0)  \mathcal P_{\underset{n_0}{J}}^{j}(\bold k_0)  \mathcal P_{\underset{n_1}{K}}^{k}(\bold k_2 + \bold k_3 - \bold k_0 - \bold G)|_{\underset{k_2^x = \xi}{}}  \mathcal P_{\underset{n_2}{M}}^{m}(\bold k_2)|_{\underset{k_2^x = \xi}{}}  \mathcal P_{\underset{n_3}{N}}^{n}(\bold k_3) \textrm{ ,}
\end{split} \label{eq:kscat3} 
\end{equation}
\end{strip}

\begin{figure}
 \centering
 \includegraphics[width=0.4\textwidth]{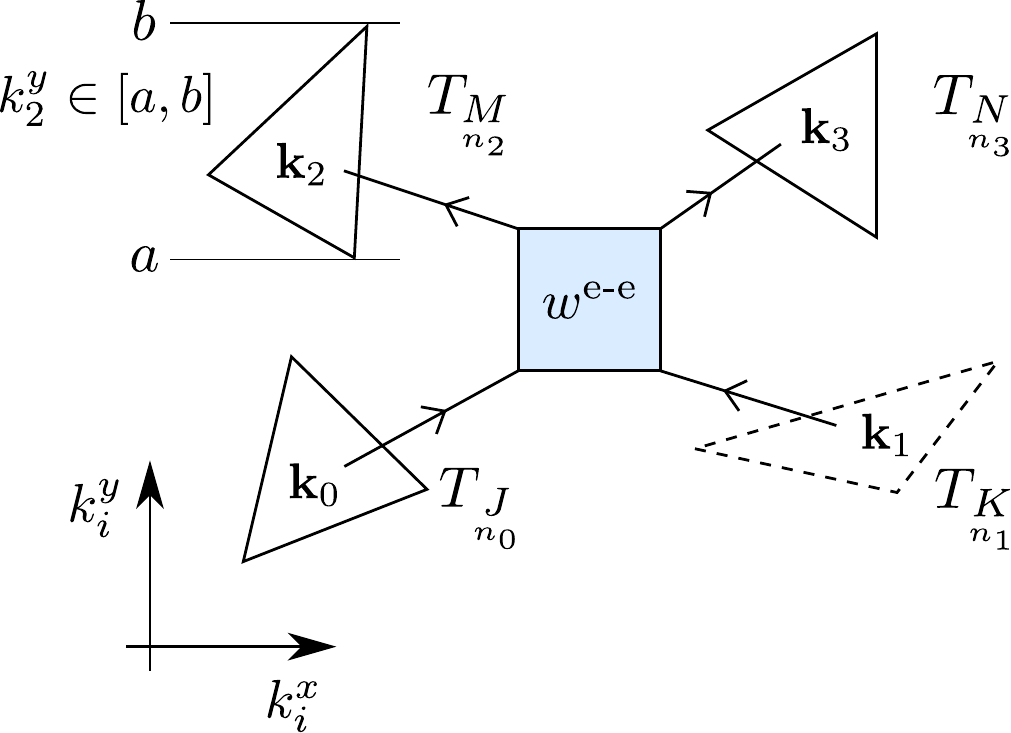}
\caption{Schematic picture of the electron-electron scattering process with the local elements (triangles). The momentum $\bold k_1$ is used for the inversion of the momentum-conserving delta (dashed triangle). The momentum $k_2^x$ is used for the inversion of the energy-conserving delta.}
  \label{fig:integration_meth}
\end{figure}

\begin{strip}
with
\end{strip}
\begin{strip}
\begin{equation} 
\begin{split}
\xi = \frac{1}{u_{\underset{n_1}{K}}^x -  u_{\underset{n_2}{M}}^x} \Bigg( t_{\underset{n_2}{M}} &+ t_{\underset{n_3}{N}} - t_{\underset{n_0}{J}} - t_{\underset{n_1}{K}} + \bold u_{\underset{n_1}{K}} \cdot \bold G - 
 \left ( \bold u_{\underset{n_0}{J}} - \bold u_{\underset{n_1}{K}}  \right ) \cdot \bold k_0 - \left ( u_{\underset{n_1}{K}}^y - u_{\underset{n_2}{M}}^y  \right )  k_2^y  - \left ( \bold u_{\underset{n_1}{K}} - \bold u_{\underset{n_3}{N}}  \right ) \cdot \bold k_3 \Bigg ) \textrm{ .}
\end{split}
\end{equation}
\end{strip}
In Eq.~\eqref{eq:kscat3} the new integration domain is $T_{\underset{n_0}{J}} \times T_{\underset{n_3}{N}} \times L_{\underset{n_2}{M}}^{[a,b]}$ where $L_{\underset{n_2}{M}}^{[a,b]}$ is a line interval from $a$ to $b$, where $a$ ($b$) is the minimum (maximum) $k_2^y$-value in the corresponding triangle (Fig.~\ref{fig:integration_meth}). Additionally, another $\Omega$-function occurs in Eq.~\eqref{eq:kscat3} that ensures that the $\bold k_2$ momentum stays inside its triangle. 

The integral Eq.~\eqref{eq:kscat3} contains a smooth integrand (except for the $\Omega$-functions) which can be computed numerically with Monte Carlo methods. For that purpose we generate a number $N_\textrm{MC}$ of sets of random points $\bold k_{\underset{\alpha}{0}} \in T_{\underset{n_0}{J}}$, $\bold k_{\underset{\alpha}{3}} \in T_{\underset{n_3}{N}} $ and $k_{\underset{\alpha}{2}}^y \in  L_{\underset{n_2}{M}}^{[a,b]}$. The scattering tensor is then calculated according to

\begin{strip}
\begin{equation}
\begin{split}
 &\left ( \mathbb S _{n_0 + n_1 \leftrightarrow n_2 + n_3 } \right )^{ijkmn}_{\underset{n_0 n_1 n_2 n_3}{IJKMN}} =  \delta_{I,J} \frac{T_{\underset{n_0}{J}} \times T_{\underset{n_3}{N}} \times L_{\underset{n_2}{M}}^{[a,b]}}{N_\textrm{MC} \big | u_{\underset{n_1}{K}}^x -  u_{\underset{n_2}{M}}^x \big |} \sum_{\bold G}  \sum_{\alpha=1}^{N_\textrm{MC}}w_{0123}^\textrm{e-e} ~\Omega( (\bold k_{\underset{\alpha}{2}} + \bold k_{\underset{\alpha}{3}} - \bold k_{\underset{\alpha}{0}} - \bold G) \in T_{\underset{n_1}{K}} )|_{\underset{k_2^x = \xi_\alpha}{}} ~ \Omega(\bold k_{\underset{\alpha}{2}} \in T_{\underset{n_2}{M}} )|_{\underset{k_2^x = \xi_\alpha}{}} \\ 
 & \quad \quad \quad \quad  \quad  \quad \quad \quad \quad  \quad \quad \quad  \quad\quad \quad \quad \quad  \times  \mathcal P_{\underset{n_0}{I}}^{i}(\bold k_{\underset{\alpha}{0}})  \mathcal P_{\underset{n_0}{J}}^{j}(\bold k_{\underset{\alpha}{0}})  \mathcal P_{\underset{n_1}{K}}^{k}(\bold k_{\underset{\alpha}{2}} + \bold k_{\underset{\alpha}{3}} - \bold k_{\underset{\alpha}{0}} - \bold G)|_{\underset{k_{\underset{\alpha}{2}}^x = \xi_\alpha}{}}  \mathcal P_{\underset{n_2}{M}}^{m}(\bold k_{\underset{\alpha}{2}})|_{\underset{k_{\underset{\alpha}{2}}^x = \xi_\alpha}{}}  \mathcal P_{\underset{n_3}{N}}^{n}(\bold k_{\underset{\alpha}{3}})  \textrm{ .}
\end{split} \label{eq:kscat4} 
\end{equation}
\end{strip}

The assumption $\left (  u_{\underset{n_1}{K}}^x -  u_{\underset{n_2}{M}}^x  \right ) \neq 0$ was necessary for the inversion of the energy-delta. This does not hold in general, hence, the algorithm has to decide which momentum component to use for the inversion. 

Our method always uses the momentum where the corresponding prefactor has the largest absolute-value. To understand why this is reasonable, we study the scattering-tensor elements where $i=j=k=m=n=0$. 
With the definition of the basis functions Eq.~\eqref{eq:dgBasis2D_2}, the Monte Carlo integration in Eq.~\eqref{eq:kscat4} becomes

\begin{strip}
\begin{equation}
\begin{split}
 \left ( \mathbb S _{n_0 + n_1 \leftrightarrow n_2 + n_3 } \right )^{00000}_{\underset{n_0 n_1 n_2 n_3}{IJKMN}} =  \delta_{I,J} \frac{T_{\underset{n_0}{J}} \times T_{\underset{n_3}{N}} \times L_{\underset{n_2}{M}}^{[a,b]}}{N_\textrm{MC} \big | u_{\underset{n_1}{K}}^x -  u_{\underset{n_2}{M}}^x \big |}& \gamma^0_{\underset{n_0}{I}} \gamma^0_{\underset{n_0}{J}} \gamma^0_{\underset{n_1}{K}} \gamma^0_{\underset{n_2}{M}} \gamma^0_{\underset{n_3}{N}} ~ w^\textrm{e-e}  \underbrace{\sum_{\bold G}  \sum_{\alpha=1}^{N_\textrm{MC}} ~\Omega(\bold k_{\underset{\alpha}{1}} \in T_{\underset{n_1}{K}} ) ~ \Omega(\bold k_{\underset{\alpha}{2}} \in T_{\underset{n_2}{M}} )|_{\underset{k_2^x = \xi}{}}}_{N_\textrm{acc}}  \textrm{ .}
\end{split} \label{eq:kscat5} 
\end{equation}
\end{strip}

\begin{figure*}[tb]
 \centering
 \includegraphics[width=0.8\textwidth]{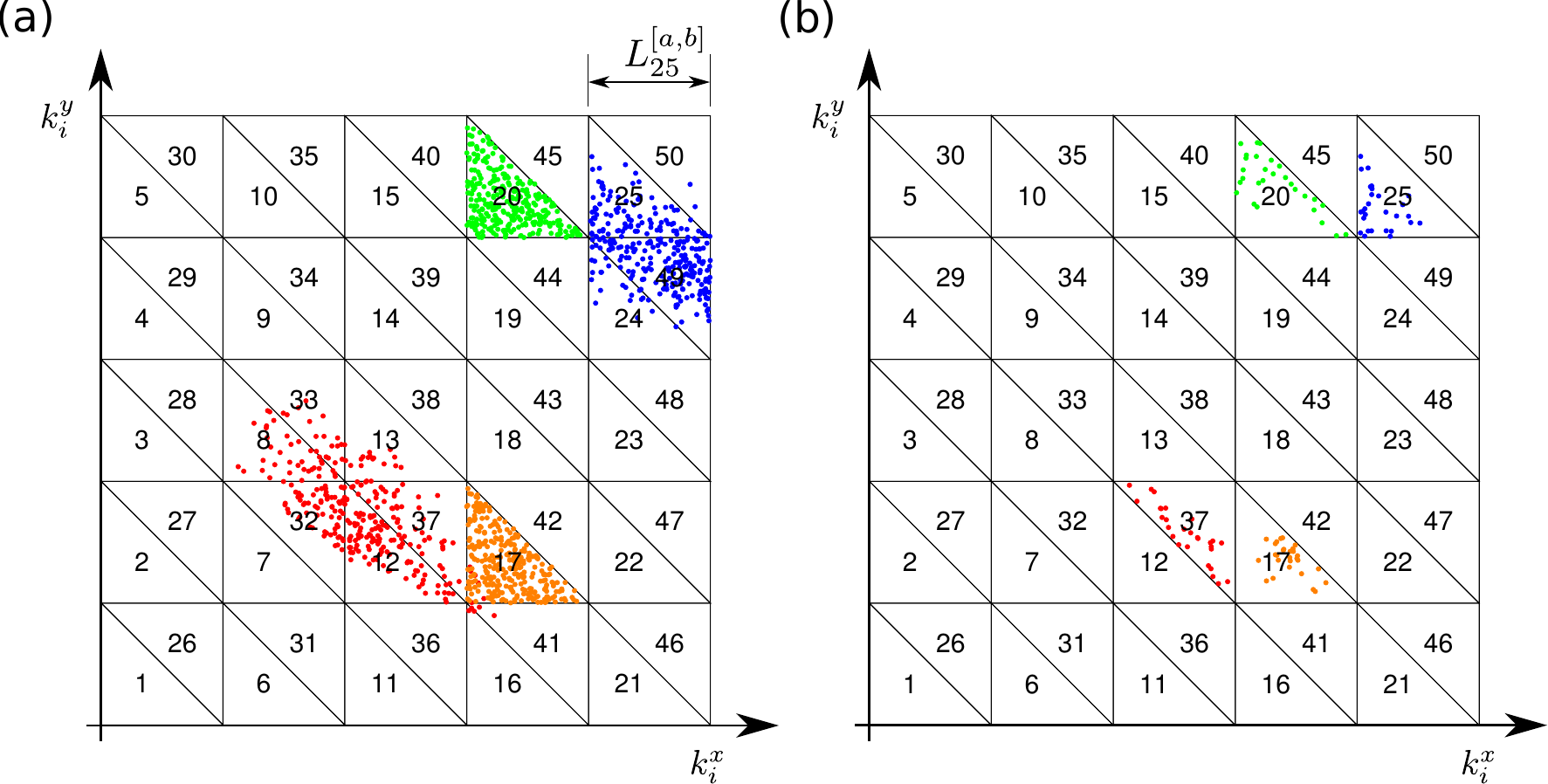}
\caption{Mesh with the element numbering on top for a $2D$-system. (a) Energy- and momentum-conserving Monte Carlo points for the scattering $n_0 + n_0 \leftrightarrow n_0 + n_0$ with the elements $J=25,K=37,M=20,N=17$ ($N_\textrm{monte} = 300$). The effective integration domain the algorithm chose is $T_{20} \times T_{17} \times L_{25}^{[a,b]}$. (b) The same point-sets as in (a) but only those where all momenta lie within their corresponding elements (i.e. the accepted Monte Carlo points $N_\textrm{acc} = 28$).}
  \label{fig:integration_meth_points}
\end{figure*}

In Eq.~\eqref{eq:kscat5} we have approximated the scattering amplitude as momentum independent, i.e. $w^\textrm{e-e}_{0123} \to w^\textrm{e-e}$ and written it in front of the momentum-sum. This is a good approximation as the momenta are restricted to their triangles and the dependence of the scattering amplitude on the momenta is usually weak.
With this approximation, the only factors that depend on the momentum that was chosen for the inversion are $L_{\underset{n_2}{M}}^{[a,b]}$, $\big | u_{\underset{n_1}{K}}^x -  u_{\underset{n_2}{M}}^x \big |$ and the number of accepted Monte Carlo points $N_\textrm{acc}$ (see Fig.~\ref{fig:integration_meth_points}). If the mesh contains only triangles that have approximately the same area and that are well conditioned (that means that all sides have approximately the same length), then the line-element $L_{\underset{n_2}{M}}^{[a,b]}$ is almost irrespective of the chosen inversion.

In general, the calculation must give the same scattering-tensor element for all possible inversions. Hence, if the factor $\big | u_{\underset{n_1}{K}}^x -  u_{\underset{n_2}{M}}^x \big |$ is largest of all possible inversions, also the number of accepted Monte Carlo points is the largest. This, in turn, makes the Monte Carlo error the smallest.

\section{More details on conservation symmetries}\label{chap:cons_sym}
\subsection{Symmetries}
As mentioned above, a problem of the Monte Carlo calculation of the tensor elements is the conservation of extensive quantities such as particle number, energy or momentum. 

The total change of an extensive quantity $\frac{\partial}{\partial t} \Theta_{\underset{n_0}{I}}$ in one element $T_{\underset{n_0}{I}}$ due to the electron-electron scattering process $n_0 + n_1 \leftrightarrow n_2 + n_3$ reads 

\begin{strip}
\begin{equation}
\begin{split}
\frac{\mathrm d }{\mathrm d t} \Theta_{\underset{n_0}{J}} &  =  \int_{T_{\underset{n_0}{J}}} \mathrm d^2k_0 ~\theta_{n_0} \left ( \frac{\partial f_0}{\partial t} \right )_{n_0 + n_1 \leftrightarrow n_2 + n_3 } =  \\
&   =  \sum_{\underset{K,M,N}{i,j,k,m,n}} \theta_{\underset{n_0}{J}}^i \left ( \mathbb S _{n_0 + n_1 \leftrightarrow n_2 + n_3 } \right )^{ijkmn}_{\underset{n_0 n_1 n_2 n_3}{JJKMN}} \Big (  ( 1_{\underset{n_0}{J}}^{j} - f_{\underset{n_0}{J}}^{j}   )   ( 1_{\underset{n_1}{K}}^{k} - f_{\underset{n_1}{K}}^{k}   ) f_{\underset{n_2}{M}}^{m}  f_{\underset{n_3}{N}}^{n}  - f_{\underset{n_0}{J}}^{j} f_{\underset{n_1}{K}}^{k}  ( 1_{\underset{n_2}{M}}^{m} - f_{\underset{n_2}{M}}^{m}   ) ( 1_{\underset{n_3}{N}}^{n} - f_{\underset{n_3}{N}}^{n} ) \Big ) \textrm{ ,}
\end{split}\label{eq:timeDerEx}
\end{equation}
\end{strip}
with the projection of the single-particle contribution $\theta_{\underset{n_0}{J}}^i$ (see Eq.~\eqref{eq:singlePartProj}). 
Here we have also used the property
\begin{equation}
\left ( \mathbb S_\textrm{col.}  \right )^{ijkmn}_{\underset{n_0 n_1 n_2 n_3}{IJKMN}} = \delta_{I,J} \left ( \mathbb S_\textrm{col.}  \right )^{ijkmn}_{\underset{n_0 n_1 n_2 n_3}{IIKMN}} \textrm{ ,} \label{eq:IJsymm}
\end{equation}
that follows from the local nature of the DG basis.
In the following we will drop the process-label and denote the scattering tensor only with $\mathbb S ^{ijkmn}_{\underset{n_0 n_1 n_2 n_3}{JJKMN}}$ for brevity. Furthermore we will assume that all the four bands that are involved in the scattering process are different (i.e. $n_0 \neq n_1 \neq n_2 \neq n_3$). For the cases where some of the bands are the same one can derive the corresponding equations in a similar way as explained below.

We can now choose a specific set of elements $J,K,M,N$ and study the change of the extensive quantity due to scattering processes only between those elements. For this it is also advisable to split the time-derivative of the extensive quantity into two different contributions: One that comes from electrons scattered into the element ($\frac{\mathrm d }{\mathrm d t} \Theta_{\underset{n_0}{J \gets}}$) and one that comes from electrons scattered away from the element ($\frac{\mathrm d }{\mathrm d t} \Theta_{\underset{n_0}{J \to}}$); $\frac{\mathrm d }{\mathrm d t} \Theta_{\underset{n_0}{J}} = \frac{\mathrm d }{\mathrm d t} \Theta_{\underset{n_0}{J \gets}} + \frac{\mathrm d }{\mathrm d t} \Theta_{\underset{n_0}{J \to}}$. 

The extensive quantity has to be conserved not only in the full scattering process where we sum the contributions of electrons scattered into the element and contributions of electrons scattered away from the element, but also for each of the processes separately, i.e.
\begin{align}
\frac{\mathrm d }{\mathrm d t} \Theta_{\underset{n_0}{J \to}} + \frac{\mathrm d }{\mathrm d t} \Theta_{\underset{n_1}{K \to}} + \frac{\mathrm d }{\mathrm d t} \Theta_{\underset{n_2}{M \gets}} + \frac{\mathrm d }{\mathrm d t} \Theta_{\underset{n_3}{N \gets}}  &= 0 \textrm{ ,} \label{eq:excons1}\\
\frac{\mathrm d }{\mathrm d t} \Theta_{\underset{n_0}{J \gets}} + \frac{\mathrm d }{\mathrm d t} \Theta_{\underset{n_1}{K \gets}} + \frac{\mathrm d }{\mathrm d t} \Theta_{\underset{n_2}{M \to}} + \frac{\mathrm d }{\mathrm d t} \Theta_{\underset{n_3}{N \to}}  &= 0 \textrm{ .}\label{eq:excons2}
\end{align}
In Eq.~\eqref{eq:timeDerEx} the part containing the projected distribution function $f_{\underset{n}{I}}^{i}$  (i.e. the phase space factor) consists of two terms which represent the scattering into the current element and away from it. Hence, if we want to calculate the conservation equations Eq.~\eqref{eq:excons1} and Eq.~\eqref{eq:excons2} we have to use only one part of the phase-space factor. 

Eq.~\eqref{eq:excons1} then becomes

\begin{strip}
\begin{equation}
\begin{split}
0=&\sum_{{i,j,k,m,n}} \Bigg [ \theta_{\underset{n_0}{J}}^i  ~\left ( \mathbb S ^{ijkmn}_{\underset{n_0 n_1 n_2 n_3}{JJKMN}} + \mathbb S ^{ijknm}_{\underset{n_0 n_1 n_3 n_2}{JJKNM}} \right )  \Big (  - f_{\underset{n_0}{J}}^{j} f_{\underset{n_1}{K}}^{k}  ( 1_{\underset{n_2}{M}}^{m} - f_{\underset{n_2}{M}}^{m}   ) ( 1_{\underset{n_3}{N}}^{n} - f_{\underset{n_3}{N}}^{n} ) \Big )\\
&+ \theta_{\underset{n_1}{K}}^i  ~\left ( \mathbb S ^{ikjmn}_{\underset{n_1 n_0 n_2 n_3}{KKJMN}} + \mathbb S ^{ikjnm}_{\underset{n_1 n_0 n_3 n_2}{KKJNM}} \right )  \Big (  - f_{\underset{n_1}{K}}^{k} f_{\underset{n_0}{J}}^{j}  ( 1_{\underset{n_2}{M}}^{m} - f_{\underset{n_2}{M}}^{m}   ) ( 1_{\underset{n_3}{N}}^{n} - f_{\underset{n_3}{N}}^{n} ) \Big )\\
&+ \theta_{\underset{n_2}{M}}^i  ~\left ( \mathbb S ^{imnjk}_{\underset{n_2 n_3 n_0 n_1}{MMNJK}} + \mathbb S ^{imnkj}_{\underset{n_2 n_3 n_1 n_0}{MMNKJ}} \right )  \Big (  ( 1_{\underset{n_2}{M}}^{m} - f_{\underset{n_2}{M}}^{m}   ) ( 1_{\underset{n_3}{N}}^{n} - f_{\underset{n_3}{N}}^{n} ) f_{\underset{n_0}{J}}^{j} f_{\underset{n_1}{K}}^{k}   \Big )\\
&+ \theta_{\underset{n_3}{N}}^i  ~\left ( \mathbb S ^{inmjk}_{\underset{n_3 n_2 n_0 n_1}{NNMJK}} + \mathbb S ^{inmkj}_{\underset{n_3 n_2 n_1 n_0}{NNMKJ}} \right )  \Big (  ( 1_{\underset{n_3}{N}}^{n} - f_{\underset{n_3}{N}}^{n} ) ( 1_{\underset{n_2}{M}}^{m} - f_{\underset{n_2}{M}}^{m}   )  f_{\underset{n_0}{J}}^{j} f_{\underset{n_1}{K}}^{k}  \Big ) \Bigg ] \textrm{ ,}
\end{split} \label{eq:exconsLong}
\end{equation}
\end{strip}
where we have only taken into account the scatterings between the elements $J,K,M,N$. All the partial phase-space factors occurring in Eq.~\eqref{eq:exconsLong} are the same (except for the signs). This partial phase-space factor consists of several terms of different powers of the distribution function, 
\begin{strip}
\begin{equation}
f_{\underset{n_0}{J}}^{j} f_{\underset{n_1}{K}}^{k}  ( 1_{\underset{n_2}{M}}^{m} - f_{\underset{n_2}{M}}^{m}   ) ( 1_{\underset{n_3}{N}}^{n} - f_{\underset{n_3}{N}}^{n}) = \underbrace{f_{\underset{n_0}{J}}^{j} f_{\underset{n_1}{K}}^{k}  1_{\underset{n_2}{M}}^{m} 1_{\underset{n_3}{N}}^{n}}_{\propto ~  f^2} -\underbrace{ f_{\underset{n_0}{J}}^{j} f_{\underset{n_1}{K}}^{k}  f_{\underset{n_2}{M}}^{m} 1_{\underset{n_3}{N}}^{n}  - f_{\underset{n_0}{J}}^{j} f_{\underset{n_1}{K}}^{k}  1_{\underset{n_2}{M}}^{m} f_{\underset{n_3}{N}}^{n}}_{\propto ~ f^3} + \underbrace{f_{\underset{n_0}{J}}^{j} f_{\underset{n_1}{K}}^{k}  f_{\underset{n_2}{M}}^{m} f_{\underset{n_3}{N}}^{n} }_{\propto~ f^4} \textrm{ .} \label{eq:partialPhase}
\end{equation}
\end{strip}
Eq.~\eqref{eq:exconsLong} must hold for each of these terms separately as they scale with different powers of the distribution function vector $f_{\underset{n}{I}}^{i}$. 

The distribution-function may have arbitrary shapes within every element, the conservation equations must nevertheless hold. Hence, we may apply the variational principle for the projected distribution-function components. For the term $\propto ~ f^4$ this gives the following conservation equation for the quantity $\Theta_n$,

\begin{strip}
\begin{equation}
\begin{split}
0=& \sum_i \Bigg [\theta_{\underset{n_0}{J}}^i  ~\left ( \mathbb S ^{ijkmn}_{\underset{n_0 n_1 n_2 n_3}{JJKMN}} + \mathbb S ^{ijknm}_{\underset{n_0 n_1 n_3 n_2}{JJKNM}} \right ) + \theta_{\underset{n_1}{K}}^i  ~\left ( \mathbb S ^{ikjmn}_{\underset{n_1 n_0 n_2 n_3}{KKJMN}} + \mathbb S ^{ikjnm}_{\underset{n_1 n_0 n_3 n_2}{KKJNM}} \right )\\
&- \theta_{\underset{n_2}{M}}^i  ~\left ( \mathbb S ^{imnjk}_{\underset{n_2 n_3 n_0 n_1}{MMNJK}} + \mathbb S ^{imnkj}_{\underset{n_2 n_3 n_1 n_0}{MMNKJ}} \right ) -  \theta_{\underset{n_3}{N}}^i  ~\left ( \mathbb S ^{inmjk}_{\underset{n_3 n_2 n_0 n_1}{NNMJK}} + \mathbb S ^{inmkj}_{\underset{n_3 n_2 n_1 n_0}{NNMKJ}} \right )\Bigg] \textrm{ .}
\end{split} \label{eq:exconsLong2}
\end{equation}
\end{strip}
We have now obtained a relation between several tensor elements that is independent of the distributions functions. When the above equation holds, all the other equations that stem from the lower-order contributions in Eq.~\eqref{eq:exconsLong} (i.e. the terms that are $\propto~f^3$ and $\propto~f^2$) are automatically fulfilled as well. Notice that if we derive everything above starting from Eq.~\eqref{eq:excons2} instead of  Eq.~\eqref{eq:excons1} we arrive at exactly the same relation.   

As long as Eq.~\eqref{eq:exconsLong2} is fulfilled, the extensive quantity $\Theta$ is conserved exactly in the scattering process. It is interesting to note that Eq.~\eqref{eq:exconsLong2} couples only tensor elements which are permutations of the index triples $\{J,j,n_0\},\{K,k,n_1\},\{M,m,n_2\},\{N,n,n_3\}$. In a $2D$-system with basis functions up to linear order (i.e. three basis functions per element) this means that 24 tensor elements are coupled by the conservation equations. Furthermore, the conservation equations do not couple different groups of these 24 tensor elements which makes them independent subspaces when it comes to the enforcement of the symmetries.

There are more symmetries in the tensors than the symmetries that stem from conserved quantities. First, it is easy to see from the definition of the tensor that
\begin{equation}
\mathbb S ^{ijkmn}_{\underset{n_0 n_1 n_2 n_3}{JJKMN}} =  \mathbb S ^{ijknm}_{\underset{n_0 n_1 n_3 n_2}{JJKNM}} \textrm{ ,} \label{eq:symmlasttwo}
\end{equation}
i.e. there is symmetry in the last two index triples. Note, that this symmetry depends on the specific process under consideration. For example a four-leg process where three legs go into the vertex and only one comes out does not have this symmetry in general. 

Furthermore, the tensors possess symmetry in the first two, small indices, 
\begin{equation}
 \mathbb S ^{ijkmn}_{\underset{n_0 n_1 n_2 n_3}{JJKMN}} =  \mathbb S ^{jikmn}_{\underset{n_0 n_1 n_2 n_3}{JJKMN}} \textrm{ ,} \label{eq:symmIJ}
\end{equation}
which holds independently of the specific scattering process described by the tensor. 
As already explained above, the conservation symmetries Eq.~\eqref{eq:exconsLong2} and also the symmetry in the last two index-triples Eq.~\eqref{eq:symmlasttwo} couple only 24 tensor elements in a $2D$-system. However, the symmetry Eq.~\eqref{eq:symmIJ} couples these different groups of 24 tensor-elements. Hence, taking all symmetries together, in a $2D$-system a number of $N_\textrm{couple}=3^5 \times 8 = 1944$ is coupled by the symmetries. Still, the symmetries can be restored in these blocks of $N_\textrm{couple}$ independently. 

One more symmetry follows from the definition of the scattering tensors and the basis Eq.~\eqref{eq:dgBasis2D_2}. As the basis functions are defined to be orthonormal, using Eq.~\eqref{eq:oneProj} leads to
\begin{equation}
 \Phi_{\underset{n}{I}}^{0}(\bold k) =  \gamma_{\underset{n}{I}}^{0} \overset{!}{=} \frac{1}{ 1_{\underset{n}{I}}^{0}} \textrm{ .}
\end{equation}
With this relation it is easy to show that
\begin{equation}
\begin{split}
&1_{\underset{n_0}{J}}^{0} ~  \mathbb S ^{0jkmn}_{\underset{n_0 n_1 n_2 n_3}{JJKMN}} = 
1_{\underset{n_1}{K}}^{0} ~  \mathbb S ^{0kjmn}_{\underset{n_1 n_0 n_2 n_3}{KKJMN}} = \\
&=1_{\underset{n_2}{M}}^{0} ~  \mathbb S ^{0mnjk}_{\underset{n_2 n_3 n_0 n_1}{MMNJK}} = 
1_{\underset{n_3}{N}}^{0} ~  \mathbb S ^{0nmjk}_{\underset{n_3 n_2 n_0 n_1}{NNMJK}} \textrm{ .} \label{eq:symm}
\end{split}
\end{equation}
Interestingly, if the tensors fulfill the symmetries Eq.~\eqref{eq:symm} and Eq.~\eqref{eq:symmlasttwo}, particle conservation in the scattering process is automatically ensured. This can be easily proven when the above symmetries are inserted in the equation for particle-conservation (Eq.~\eqref{eq:exconsLong2} with $\theta_{\underset{n_0}{J}}^i = 1_{\underset{n_0}{J}}^i$)

\begin{strip}
\begin{equation}
\begin{split}
0=&  \Bigg [1_{\underset{n_0}{J}}^0  ~\left ( \mathbb S ^{0jkmn}_{\underset{n_0 n_1 n_2 n_3}{JJKMN}} + \mathbb S ^{0jknm}_{\underset{n_0 n_1 n_3 n_2}{JJKNM}} \right ) + 1_{\underset{n_1}{K}}^0  ~\left ( \mathbb S ^{0kjmn}_{\underset{n_1 n_0 n_2 n_3}{KKJMN}} + \mathbb S ^{0kjnm}_{\underset{n_1 n_0 n_3 n_2}{KKJNM}} \right )\\
&- 1_{\underset{n_2}{M}}^0  ~\left ( \mathbb S ^{0mnjk}_{\underset{n_2 n_3 n_0 n_1}{MMNJK}} + \mathbb S ^{0mnkj}_{\underset{n_2 n_3 n_1 n_0}{MMNKJ}} \right ) -  1_{\underset{n_3}{N}}^0  ~\left ( \mathbb S ^{0nmjk}_{\underset{n_3 n_2 n_0 n_1}{NNMJK}} + \mathbb S ^{0nmkj}_{\underset{n_3 n_2 n_1 n_0}{NNMKJ}} \right )\Bigg] \textrm{ ,}
\end{split} \label{eq:exconsLong3}
\end{equation}
\end{strip}
where we have used the property $1_{\underset{n_0}{J}}^i = \delta_{i,0} 1_{\underset{n_0}{J}}^0$.

\section{Enforce the symmetries} \label{chapA:cleanup_details}
As long as the equations ensuing from the symmetries that correspond to physically conserved quantities are fulfilled exactly, the numerical scattering conserves these quantities exactly as well. However, as explained in section~\ref{chap:calcSC} we use a Monte Carlo technique to calculate the tensors and therefore the tensor-elements are subject to a finite Monte Carlo error. Hence, we need to restore the symmetries in the tensor. 

One possibility would be to calculate only some of the tensor-elements using the Monte Carlo routine and determine the remaining elements from the symmetry equations of the previous section. This brings the problem that we need a strategy to determine the best inversion of the equations. This can be done and was, indeed, tested. In the following we will explain another method that proved to be more stable and easier to implement.

The symmetry-equations derived in the previous section are all linear equations in the tensor-elements. Therefore we can think of these equations (Eq.~\eqref{eq:exconsLong2}, \eqref{eq:symmlasttwo}, \eqref{eq:symmIJ}, \eqref{eq:symm}) as a scalar product of two vectors, the first vector represents the $i$-th equation ($\bold v_i \quad i \in [1,N_\textrm{sym}]$ where $N_\textrm{sym}$ is the number of symmetry-equations), the second one is the vector representation of the involved tensor-elements ($\mathbb S_v$). The symmetry-equation is fulfilled if the scalar product of those two vectors vanishes (i.e. $\bold v_i \cdot \mathbb S_v = 0$). The dimensionality of the space in which these vectors live is equal to the number of coupled tensor-elements $N_\textrm{couple}$ introduced in the previous section.

As an example we take one of the symmetry-equations for the last two index-triples (Eq.~\eqref{eq:symmlasttwo}) for the coupled tensor-elements of the element-combination $\{J,n_0\}$, $\{K,n_1\}$, $\{M,n_2\}$, $\{N,n_3\}$. The corresponding vectors are 
\begin{equation}
\bold v_1 = \begin{pmatrix}
 1\\
 -1\\
 0\\
 \vdots \\
 0
\end{pmatrix}
\textrm{ , } \quad
\mathbb S_v = \begin{pmatrix}
 \mathbb S ^{0 0000}_{\underset{n_0 n_1 n_2 n_3}{JJKMN}}\\
 \mathbb S ^{0 0000}_{\underset{n_0 n_1 n_3 n_2}{JJKNM}}\\
 \mathbb S ^{10000}_{\underset{n_0 n_1 n_2 n_3}{JJKMN}}\\
 \vdots \\
 \mathbb S ^{22222}_{\underset{n_3 n_2 n_1 n_0}{NNMKJ}}
\end{pmatrix}
\end{equation}
for which the symmetry equation is obtained if the scalar-product is zero,
\begin{equation}
0\overset{!}{=} \bold v_1 \cdot \mathbb S_v = \mathbb S ^{00000}_{\underset{n_0 n_1 n_2 n_3}{JJKMN}} - \mathbb S ^{00000}_{\underset{n_0 n_1 n_3 n_2}{JJKNM}}  \to   \mathbb S ^{00000}_{\underset{n_0 n_1 n_2 n_3}{JJKMN}} = \mathbb S ^{00000}_{\underset{n_0 n_1 n_3 n_2}{JJKNM}} \textrm{ .} 
\end{equation}

The equation-vectors $\bold v_i$ together with $\mathbb S_v$ form a set of vectors that spans a subspace of the the vector-space they live in. We define this set as,  
\begin{equation} 
 \bold \xi_i= \left\{
\begin{array}{ll}
     \bold v_i & \quad \textrm{if } i \in [1,N_\textrm{sym}] \\ 
    \mathbb S_v & \quad \textrm{if } i=N_\textrm{sym}+1
\end{array} \right. .
\end{equation}  
This set of vectors spans a $(N_\textrm{sym}+1)$-dimensional space and consists of non-orthogonal vectors in general. A fully orthogonal set of vectors $ \chi _i$ that spans the same space can be obtained by the so-called Gram-Schmidt procedure.

According to the Gram-Schmidt procedure, the orthogonalized vectors are obtained by,
\begin{equation}
 \chi_1 = \xi_1
\end{equation}
and 
\begin{equation}
 \chi_i = \xi_i -  \sum_{j=1}^{i-1} \frac{\xi_i \cdot \chi_j}{\chi_j \cdot \chi_j} \chi_j \quad \forall i \in [2,N_\textrm{sym}+1] \textrm{ .}
\end{equation}
The last one of the orthogonalized vector set is the sought tensor within the Monte Carlo error-sphere that fulfils all symmetries exactly , i.e.
\begin{equation}
\tilde{\mathbb S}_v = \chi_{N_\textrm{sym}+1}\textrm{ .}
\end{equation}


\section{Scattering rates}\label{chap:Ascatteringrates}

In the following we will refer to scattering rates of fermions, yet the description applies with minimal adjustments to different types of scatterings.

For that purpose we study a distribution-function that consists of a Fermi-Dirac distribution $f_\textrm{FD}(\epsilon_n(\bold k),\mu,\beta)$ plus an excitation $\delta f_n(\bold k)$,
\begin{equation}
f_n(t, \bold k) = f_\textrm{FD}(\epsilon_n(\bold k),\mu,\beta) +  \delta f_n(t, \bold k) \textrm{ .} \label{eq:fDistortion}
\end{equation} 
with the excitation $\delta f_{n}(t, \bold k)$  small in amplitude and localized in momentum space (i.e. a single electron/hole added at a certain momentum $\bold k_0$).

When we insert Eq.~\eqref{eq:fDistortion} into Eq.~\eqref{eq:eecollisionScatNum} we get 

\begin{strip}
\begin{equation}
\begin{split}
 \frac{\partial (\delta f_0)}{\partial t}  & = \sum_{\bold G}  \iiint   \mathrm d^d k_1 \mathrm d^d k_2 \mathrm d^d k_3   ~\delta_{\bold k}  \delta_{\epsilon} \\
&\times   \Big [ (1-f_{\textrm{FD}_0})(1-f_{\textrm{FD}_1})f_{\textrm{FD}_2} f_{\textrm{FD}_3} - f_{\textrm{FD}_0} f_{\textrm{FD}_1} (1-f_{\textrm{FD}_2})(1-f_{\textrm{FD}_3}) \\
&\quad \quad -\delta f_0 \Big( (1-f_{\textrm{FD}_1})f_{\textrm{FD}_2} f_{\textrm{FD}_3} + f_{\textrm{FD}_1} (1-f_{\textrm{FD}_2})(1-f_{\textrm{FD}_3} ) \Big ) \\
& \quad \quad -\delta f_1 \Big ( (1-f_{\textrm{FD}_0}) f_{\textrm{FD}_2} f_{\textrm{FD}_3} + f_{\textrm{FD}_0}  (1-f_{\textrm{FD}_2})(1-f_{\textrm{FD}_3})\Big ) \\
& \quad \quad +\delta f_2 \Big ((1-f_{\textrm{FD}_0})(1-f_{\textrm{FD}_1}) f_{\textrm{FD}_3} + f_{\textrm{FD}_0} f_{\textrm{FD}_1} (1-f_{\textrm{FD}_3}) \Big )\\
& \quad \quad +\delta f_3 \Big ( (1-f_{\textrm{FD}_0})(1-f_{\textrm{FD}_1})f_{\textrm{FD}_2} + f_{\textrm{FD}_0} f_{\textrm{FD}_1} (1-f_{\textrm{FD}_2})) \Big )\\
&\quad \quad +~ \cdots
 \label{eq:eecollisionScatRate2}
\end{split}  \textrm{ ,}
\end{equation}
\end{strip}
where we have omitted all but the leading order terms in $\delta f$. The second line in Eq.~\eqref{eq:eecollisionScatRate2} is just the electron-electron phase-space factor evaluated with Fermi-Dirac distributions. The Fermi-Dirac distribution is a fixed point of the collision integral, hence, the integral of this term equals zero. The term in the third line is $\propto \delta f(\bold k_0)$. The momentum $\bold k_0$ is not an integration variable of the integrals and therefore the term $\delta f(\bold k_0)$ can be written in front of the integrals. With all the remaining terms grouped together in the function $\mathcal R_0$ we can write Eq.~\eqref{eq:eecollisionScatRate2} as
\begin{equation}
 \frac{\partial (\delta f_0)}{\partial t}   =~-\delta f_0 ~ \lambda_0 + \mathcal R_0  \label{eq:eecollisionScatRate3} \textrm{ ,}
\end{equation}
with

\begin{strip}
\begin{equation}
\lambda_0 \equiv   \sum_{\bold G}  \iiint   \mathrm d^d k_1 \mathrm d^d k_2 \mathrm d^d k_3   ~ w_{0123}^\textrm{e-e} ~\delta_{\bold k}  \delta_{\epsilon} \Big( (1-f_{\textrm{FD}_1})f_{\textrm{FD}_2}f_{\textrm{FD}_3} + f_{\textrm{FD}_1} (1-f_{\textrm{FD}_2})(1-f_{\textrm{FD}_3} ) \Big )\textrm{ .} \label{eq:eecollisionScatRate4}
\end{equation}
\end{strip}
Eq.~\eqref{eq:eecollisionScatRate3} describes an exponentially decaying excitation $\delta f_0$ if the term $\mathcal R_0$ is negligible. This is indeed the case when an excitation $\delta f$ is used that is localized at $\bold k_0$ in momentum space and has a sufficiently small amplitude. This is equivalent to adding (or removing) a single particle at a momentum $\bold k_0$. Then $\lambda(\bold k_0)$ as defined by Eq.~\eqref{eq:eecollisionScatRate4} can be interpreted as the scattering rate of this particle (or hole). Interestingly, for the scattering rate it does not make a difference if we add or remove a particle as the scattering rate is irrespective of the shape of the excitation $\delta f$.

We obtain the discretized version of the scattering rate by projection onto the basis functions, 

$\lambda_{\underset{n_0}{I}}^i \equiv \int \mathrm d ^d k \lambda_{n_0}(\bold k) \Phi_{\underset{n_0}{I}}^i(\bold k)$. Together with the discretized versions of the Fermi-Dirac distributions $[f_\textrm{FD}]_{\underset{n_0}{I}}^i$ the scattering rate becomes

\begin{strip}
\begin{equation}
\begin{split}
\lambda_{\underset{n_0}{I}}^i =  & \sum_{\underset{K,M,N}{k,m,n}} \sum_{\bold G}  \iiiint   \mathrm d^d k_0  \mathrm d^d k_1 \mathrm d^d k_2 \mathrm d^d k_3   ~w_{0123}^\textrm{e-e} ~\delta_{\bold k} ~ \delta_{\epsilon} ~ \Phi_{\underset{n_0}{I}}^i(\bold k_0) \Phi_{\underset{n_1}{K}}^k(\bold k_1) \Phi_{\underset{n_2}{M}}^m(\bold k_2) \Phi_{\underset{n_3}{N}}^n(\bold k_3) \\
&\quad \quad \quad\times \Big ( (1_{\underset{n_1}{K}}^k - [f_\textrm{FD}]_{\underset{n_1}{K}}^k ) [f_\textrm{FD}]_{\underset{n_2}{M}}^m [f_\textrm{FD}]_{\underset{n_3}{N}}^n + [f_\textrm{FD}]_{\underset{n_1}{K}}^k (1_{\underset{n_2}{M}}^m - [f_\textrm{FD}]_{\underset{n_2}{M}}^m) (1_{\underset{n_3}{N}}^n - [f_\textrm{FD}]_{\underset{n_3}{N}}^n)   \Big ) \label{eq:eecollisionScatRate5}
\end{split}
\end{equation}
\end{strip}
which almost has the shape of a scattering tensor that is contracted with projections of the Fermi-Dirac distributions. In order to bring Eq.~\eqref{eq:eecollisionScatRate5} to a form that involves the scattering tensor we exploit the fact that we may write the number one as $1 = \sum_J^j 1_{\underset{n_0}{J}}^j \Phi_{\underset{n_0}{J}}^j(\bold k_0)$. When we insert this relation into Eq.~\eqref{eq:eecollisionScatRate5} we obtain

\begin{strip}
\begin{equation}
\begin{split}
\lambda_{\underset{n_0}{I}}^i =& \sum_{\underset{J,K,M,N}{j,k,m,n}} \sum_{\bold G}  \iiiint \mathrm d^d k_0  \mathrm d^d k_1 \mathrm d^d k_2 \mathrm d^d k_3   ~w_{0123}^\textrm{e-e} ~\delta_{\bold k} ~ \delta_{\epsilon} ~ \Phi_{\underset{n_0}{I}}^i(\bold k_0) \Phi_{\underset{n_0}{J}}^j(\bold k_0) \Phi_{\underset{n_1}{K}}^k(\bold k_1) \Phi_{\underset{n_2}{M}}^m(\bold k_2) \Phi_{\underset{n_3}{N}}^n(\bold k_3) \\
& \quad \quad \quad \times  \Big ( 1_{\underset{n_0}{J}}^j  \big (1_{\underset{n_1}{K}}^k - [f_\textrm{FD}]_{\underset{n_1}{K}}^k  \big ) [f_\textrm{FD}]_{\underset{n_2}{M}}^m [f_\textrm{FD}]_{\underset{n_3}{N}}^n + 1_{\underset{n_0}{J}}^j [f_\textrm{FD}]_{\underset{n_1}{K}}^k \big (1_{\underset{n_2}{M}}^m - [f_\textrm{FD}]_{\underset{n_2}{M}}^m \big ) \big (1_{\underset{n_3}{N}}^n - [f_\textrm{FD}]_{\underset{n_3}{N}}^n \big )   \Big )\\
=& \sum_{\underset{J,K,M,N}{j,k,m,n}} \mathbb S ^{ijkmn}_{\underset{n_0 n_1 n_2 n_3}{IJKMN}} 
  \Big ( 1_{\underset{n_0}{J}}^j  \big (1_{\underset{n_1}{K}}^k - [f_\textrm{FD}]_{\underset{n_1}{K}}^k  \big ) [f_\textrm{FD}]_{\underset{n_2}{M}}^m [f_\textrm{FD}]_{\underset{n_3}{N}}^n + 1_{\underset{n_0}{J}}^j [f_\textrm{FD}]_{\underset{n_1}{K}}^k \big (1_{\underset{n_2}{M}}^m - [f_\textrm{FD}]_{\underset{n_2}{M}}^m \big ) \big (1_{\underset{n_3}{N}}^n - [f_\textrm{FD}]_{\underset{n_3}{N}}^n \big )   \Big )
\end{split} \label{eq:eecollisionScatRate6}
\end{equation}
\end{strip}
which is the contraction of the scattering tensor with a certain phase-space factor evaluated from the equilibrium distribution. 

\bibliography{./main}



\end{document}